\documentclass[aps,physrev,reprint,superscriptaddress,nofootinbib,floatfix, preprintnumbers]{revtex4-2}

\usepackage{amsmath,amsfonts,amssymb}
\usepackage{bm}
\usepackage{graphicx}
\usepackage{multirow}
\usepackage{float}
\usepackage[dvipsnames]{xcolor}
\usepackage[colorlinks=true,
            linkcolor=blue,
            urlcolor=blue,
            citecolor=gray]{hyperref}

\newcommand{\SNRtot}{\textrm{SNR}_{\textrm{tot}}}
\newcommand{\SNRmem}{\textrm{SNR}_{\textrm{mem}}}

\newcommand{\orcid}[1]{\href{https://orcid.org/#1}{\includegraphics[width=8pt]{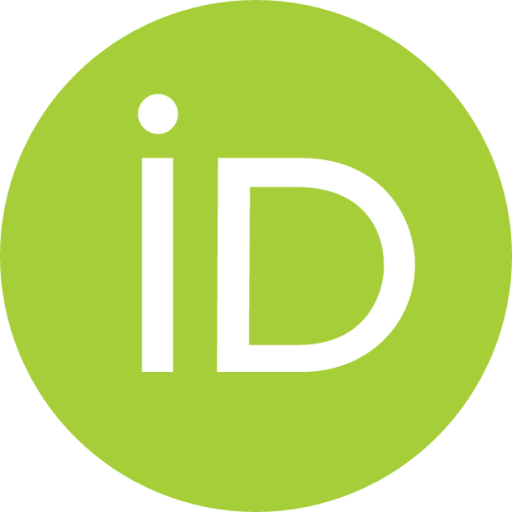}}}
\begin{document}

\preprint{CERN-TH-2026-043}
\title{Detectability of Gravitational-Wave Memory with LISA: A Bayesian Approach}

\author{Adrien Cogez \orcid{0009-0007-1316-9648}}
\email[Contact author: ]{adrien.cogez@cea.fr}
\affiliation{IRFU, CEA, Université Paris-Saclay, 91191, Gif-sur-Yvette, France}
\affiliation{Centre national d’études spatiales (CNES), Paris, France}

\author{Silvia Gasparotto~\orcid{0000-0001-7586-1786}}
\email[Contact author: ]{silvia.gasparotto@cern.ch}
\affiliation{CERN, Theoretical Physics Department, Esplanade des Particules 1, Geneva 1211, Switzerland}
\affiliation{Institut de F\'isica d’Altes Energies (IFAE), The Barcelona Institute of Science and Technology, Campus UAB, 08193 Bellaterra (Barcelona), Spain}
\author{Jann Zosso \orcid{0000-0002-2671-7531}}
\email[Contact author: ]{jann.zosso@nbi.ku.dk}
\affiliation{Center of Gravity, Niels Bohr Institute, Blegdamsvej 17, 2100 Copenhagen, Denmark}

\author{Henri Inchauspé \orcid{0000-0002-4664-6451}}
\affiliation{Institute for Theoretical Physics, KU Leuven, Celestijnenlaan 200D, B-3001 Leuven, Belgium}
\affiliation{Leuven Gravity Institute, KU Leuven, Celestijnenlaan 200D box 2415, 3001 Leuven, Belgium}

\author{Chantal Pitte~\orcid{0009-0009-0524-7292}}
\affiliation{SISSA, Via Bonomea 265, 34136 Trieste, Italy \&  INFN Sezione di Trieste}
\affiliation{IFPU - Institute for Fundamental Physics of the Universe, Via Beirut 2, 34014 Trieste, Italy}

\author{Lorena \surname{Maga\~na Zertuche} \orcid{0000-0003-1888-9904}}
\affiliation{Center of Gravity, Niels Bohr Institute, Blegdamsvej 17, 2100 Copenhagen, Denmark}

\author{Antoine Petiteau \orcid{0000-0002-7371-9695}}
\author{Marc Besancon \orcid{0000-0003-3278-3671}}
\affiliation{IRFU, CEA, Université Paris-Saclay, 91191, Gif-sur-Yvette, France}

\date{\today}

\begin{abstract}
    Gravitational wave (GW) astronomy opens a new venue to explore the universe. Future observatories such as LISA, the Laser Interferometer Space Antenna, are expected to observe previously undetectable fundamental physics effects in signals predicted by General Relativity (GR). 
    One particularly interesting such signal is associated to the displacement memory effect, which corresponds to a permanent deformation of spacetime due to the passage of gravitational radiation. 
    In this work, we explore the ability of LISA to observe and characterize this effect. In order to do this, we use state-of-the-art simulations of the LISA instrument, and we perform a Bayesian analysis to assess the detectability and establish general conditions to claim detection of the displacement memory effect from individual massive black hole binary (MBHB) merger events in LISA. We perform parameter estimation both to explore the impact of the displacement memory effect and to reconstruct its amplitude. We discuss the precision at which such a reconstruction can be obtained thus opening the way to tests of GR and alternative theories. To provide astrophysical context, we apply our analysis to black hole binary populations models and estimate the rates at which the displacement memory effect could be observed within the LISA planned lifetime.
\end{abstract}

\maketitle

\newpage
\tableofcontents
\newpage 

\section*{Introduction}
\label{sec:Introduction}
Predicted in 1916 by Einstein~\cite{AE_ondeGrav1,AE_ondeGrav2}, gravitational waves (GWs) were (directly) detected for the first time in 2015 by the LIGO-Virgo-KAGRA (LVK) Collaboration~\cite{LIGOScientific:2014pky, VIRGO:2014yos, KAGRA:2020tym, LIGOFirstGWave}. These small ripples of spacetime, caused by accelerated masses, slightly stretch and squeeze spacetime along their path. Their detection has opened a new observational window into the universe that no longer relies on electromagnetic radiation. The current ground-based interferometers are designed to see stellar-mass binary black holes; however, future facilities will offer a wider range of observations with detectors like the Laser Interferometer Space Antenna (LISA). LISA will be a space-borne detector, exploring the $0.1$ mHz to $1$ Hz frequency range~\cite{LISA_Redbook}. In the list of target sources, LISA is forecasted to detect GWs from massive black hole binaries (MBHB), which are expected to provide the strongest signals. The strength of these signals opens up the possibility to test previously unseen fundamental physics effects predicted by General Relativity (GR), such as the memory effect. This effect corresponds to a finite deformation of spacetime, either a stretch or a squeeze—relative to its initial state, which persists even after the passage of gravitational radiation. The gravitational memory effect can be separated into two main categories. First is the so-called linear memory effect, predicted in 1974 by Zel'dovich and Polnarev~\cite{MemoryEffectProposal} and developed by Braginskii, Grishchuk, and Thorne~\cite{Braginskii_Grishchuk_MemoryEffect, Braginsky_Thorne_GWBurstWithMemory}. This linear effect manifests in the presence of unbound matter in a radiation event, for example, for astrophysically compact hyperbolic encounters.
Second is the non-linear memory effect, predicted in 1991 by Christodoulou~\cite{NonLinearMemory, Thorne_1992}, which applies to all GW sources as it comes from the energy radiated in the form of GWs. This non-linear memory effect is also called the Christodoulou effect or the \textit{displacement memory} effect. This effect is linked to the asymptotic structure of spacetime through the symmetries of the Bondi-van der Burg-Metzner-Sachs (BMS) group~\cite{Pasterski_Strominger_2016, Strominger_2016, Nichols_2018, Flanagan_2019, Ashtekar_2020, Grant_Nichols_2022, Grant_Nichols_2023, Mitman_2024, AddingGWmemoryWithBMSBalanceLaw,Bhattacharjee:2019jaf}. These symmetries also predict other subleading memory effects, such as the spin~\cite{Pasterski_Strominger_2016} and center-of-mass memories~\cite{Nichols_2018, Bieri_Polnarev_2024}. However, we will focus on the dominant displacement memory effect, and it will be referred to as the ``memory effect" throughout this paper. A more detailed theoretical approach of the detection of the memory effect can be found in the companion paper,~\cite{Zosso:2025memory}.

Before entering the discussion on detectability with LISA, one should briefly mention the status of the search for the memory effect at current observatories. The memory effect has not yet been observed by the LVK Collaboration. Black hole binary mergers with black hole masses up to $\mathcal{O}$(100) solar masses constitute the primary source of transients in the frequency band of the LVK detector network, i.e., $10$ Hz - $10$ kHz~\cite{Abbott:2016xvh}. Considering that memory is an inherently low-frequency effect and the sensitivity of current GW detectors are restricted to a given frequency band one could expect that performing a direct search for a memory effect in LVK would be particularly challenging~\cite{ProspectsOfDetectingNLGWmemory, Lasky_Memory_Threshold}. Indeed, one of the most recent studies from Cheung et al.~\cite{Cheung_2024}, using the LVK GWTC-3~\cite{LIGOrun3}, points to a non-observation of the memory effect with the current public LVK data. Furthermore, it is found that a catalogue of $\mathcal{O}$(2000) binary black hole mergers would be needed to statistically detect the memory effect, as already inferred in Hübner et al.~\cite{ThanksForTheMemory, MemoryRemains}. This could occur during the fifth observing run, expected to start in 2027. In addition to the LVK Collaboration, the Pulsar Timing Array (PTA) community is also actively searching for a memory burst signal~\cite{Tomson:2025}. Recent searches for GW memory bursts include the Bayesian-based search by the NANOGrav PTA in their 12.5 yr dataset~\cite{NanoGrav12point5yMemory}, followed more recently by a search within their 15 yr dataset~\cite{Agazie_2025}. Although limits on memory strain amplitude as a function of both sky location and burst epoch have been derived, no evidence for such bursts has been found at present. Despite this possible statistical detection, a new generation of detectors such as LISA would be required to observe the signature of memory for individual events.

In this paper, we focus on the detectability of the memory effect with LISA through a Bayesian analysis on simulated datasets. It builds on the work from Inchauspé et al.~\cite{Henri_Memory_Paper}, which investigates the memory effect with the full time-domain response of LISA based on time-delay interferometry (TDI), including the most updated knowledge of the various noise components. 
Our first goal is to establish criteria for assessing the detectability of the memory effect with LISA, specifically by evaluating LISA’s ability to robustly identify the memory signal produced by the coalescence of a MBHB.
Then, we perform parameter estimation to monitor the impact of including the memory in waveform models and, in particular, assess the precision with which the amplitude of the memory burst can be estimated.
Finally, the detectability rate of the memory effect potentially observed by LISA will be assessed with the help of population models of MBHBs described in~\cite{Barausse_2020, Barausse_Lapi_2021}. A companion Git repository\footnote{https://gitlab.in2p3.fr/adrien.cogez/memory-effect-detectability-companion} is provided, featuring the figures and corresponding data.

The paper is organized as follows.
In Section~\ref{sec:MemLISA}, we describe the various GW waveforms and higher harmonics used in our simulations and how the memory effect is implemented. We also describe the method used to build mock datasets and compare them to templates, with and without the inclusion of the memory effect. 
In Section~\ref{sec:Detectability}, we compute the signal-to-noise ratio, introduce the Bayesian framework and the dynamical nested sampling method to compute the evidence and estimate posterior distributions. We also discuss the use of the Bayes factor for the memory effect detectability with LISA as well as the possible detectability criteria and strategies one can adopt to claim its detection. 
In Section~\ref{sec:ParamEstimation}, we perform parameter estimation and discuss the precision needed to estimate the amplitude of a memory burst.
Section~\ref{sec:Catalogs} is devoted to a discussion of the detectability rate of the memory effect by LISA using population models of MBHBs. 
Finally, we discuss our conclusions and outlook in Section~\ref{sec:Conclusion}.

\section{Memory effect in LISA}
\label{sec:MemLISA}
LISA will observe many MBHB signals with large signal-to-noise ratios (SNRs)~\cite{LISA_Redbook}. 
Given the loudness of these types of sources, subleading effects such as those from higher modes (HMs) and the memory effect (among other effects) should be taken into account. 
In a recent study~\cite{Henri_Memory_Paper}, the investigation of the SNR of the memory effect in LISA for different parts of parameter space provides a first indication of what kind of events would allow the detection of the memory effect in an individual event. 
However, to claim detectability, we need to assess the ability to disentangle the memory imprint from the main oscillatory component.

In this section, we provide a summary of a well-defined memory computation formula, as well as the gravitational waveform models, their higher harmonic content, and the implementation of the memory effect.

\subsection{Memory model}

Given a localized source of gravitational radiation, gravitational memory is defined as a constant offset in the proper distance separation $s$ between two freely-falling test masses at infinity after the burst radiation emission
\begin{equation}\label{eq:memorydef}
    s_1-s_0\neq 0\,,
\end{equation}
where $s_0$ defines the initial spatial proper distance separation and $s_1$ denotes the separation after the passage of the radiation.
In order to use the Bayes factor to quantify evidence in favor of the presence of gravitational memory compared to its non-existence, it is imperative to have well-defined waveform models both with and without memory. That is, a clear characterization of the memory signal within the total radiation is required. This is especially important in the case of memory, since the LISA response, as well as any other interferometric detector, will not have direct access to the defining net zero-frequency memory offset in Eq.~\eqref{eq:memorydef}, and the presence of memory can only be inferred by measuring the transition between two vacuum states. It is, thus, crucial to ensure that the defined memory signal can exclusively be associated to the net effect of leaving a permanent distortion of spacetime.

The definition of a memory model to be searched for in GW data therefore requires additional physical input. In the case of non-linear memory associated to a single compact binary coalescence (CBC), it is provided by the insight that gravitational memory is fundamentally distinct from the emission of oscillatory GWs in terms of its frequency content. That is, while the periodic GW signal is associated to the orbital time-scales of the binary and its higher harmonics, gravitational memory can directly be attributed to a monotonically raising time-domain signal, whose increase is directly related to the radiation-reaction scales of the event. Concretely, the dominant non-linear memory is directly associated to the emitted energy flux in the $+$ (plus) and $\times$ (cross) GW polarization modes
\begin{equation}\label{eq:EnergyFluxH}
    \frac{\mathrm{d}E}{\mathrm{d}t'\mathrm{d}\Omega'}=\frac{1}{16\pi G}\Big\langle \dot h^2_{H+} +\dot h^2_{H\times}\Big\rangle \,,
\end{equation}
with $G$ as Newton's gravitational constant and $(H)$ denoting the high-frequency nature of the primary waves associated to the relevant orbital timescales. The associated gravitational memory can then be computed via the general formula
\begin{equation}\label{eq:memoryequation}
    [h_{ij}^{\rm (mem)}]^{\rm TT}=\frac{4G}{d_{\rm L}}\int_{-\infty}^t \mathrm{d}t'\int \mathrm{d}\Omega' \frac{\mathrm{d}E}{\mathrm{d}t'\mathrm{d}\Omega'}\left[\frac{n'_in'_j}{1-n'_m n^m}\right]^{\rm TT}\,,
\end{equation}
where $d_{\rm L}$ is the luminosity distance, TT denotes a projection onto the transverse-traceless space, and $n_i(\Omega)$ defines a unit normal from the source to the detection location in the spherical coordinate direction $\Omega=\{\theta,\phi\}$. Observe that the memory effect is hereditary, meaning it depends on the entire past evolution of the system. Therefore, to compute the memory at an instant of retarded time $t$, we formally need to integrate from $- \infty$ to $t$.
Moreover, the explicit average over orbital scales $\langle ... \rangle$ in Eq.~\eqref{eq:EnergyFluxH} ensures gauge invariance of the energy content of  GWs~\cite{Isaacson_PhysRev.166.1263,Isaacson_PhysRev.166.1272,misner_gravitation_1973,Flanagan:2005yc,maggiore2008gravitational,Zosso:2025ffy} and implies the  frequency space separation of the resulting memory signal from the oscillatory waves. The companion paper~\cite{Zosso:2025memory} explicitly derives Eq.~\eqref{eq:memoryequation}\footnote{Note that compared to the companion paper, retarded time is denoted here as $t$ instead of $u$.} and provides further details on the insights of the formula. In particular, as discussed in Section~III C 4 of~\cite{Zosso:2025memory}, with the explicit scaling with respect to the luminosity distance, Eq.~\eqref{eq:memoryequation} directly incorporates an evolution of the memory signal over cosmological scales and thus implicitly assumes a dependence on redshifted, hence detector-frame, masses (see also~\cite{Tolish:2016ggo,Bieri:2017vni}). Up to the averaging, Eq.~\eqref{eq:memoryequation} is equivalent to the memory formula brought forth in the original literature~\cite{NonLinearMemory,Wiseman_memoryExpression,Thorne_1992}. 

The memory formula can significantly be simplified within a spin-weighted spherical harmonic expansion. Concretely, the complex spin-weight $-2$ scalar $h(t, d_{\rm L}, \theta, \varphi ) \equiv h_+ -i h_\times$ for both the GWs and the memory signal can be expanded in terms of spin-weighted spherical harmonics as 

\begin{eqnarray}
    h(t, d_{\mathrm{L}}, \theta, \varphi ) 
    &=& h_+ - i h_\times \nonumber \\
    &=& \frac{1}{d_{\mathrm{L}}} \sum_{\ell=2}^{\infty} \sum_{m=-\ell}^{\ell} 
        {}^{-2}Y_{\ell m}(\theta, \varphi) \, h_{\ell m}(t)\,.
    \label{eq:ModesToPolar}
\end{eqnarray}

As explicitly shown in the companion paper~\cite{Zosso:2025memory} (see also~\cite{Favata:2008yd,GWmemory}), for a quasi-circular binary black hole merger, the dominant contribution to Eq.~\eqref{eq:memoryequation} is given by the $(\ell=2,m=0)$ mode\footnote{Adding precession, will for instance, lead to additional significant components of memory in $m = \pm 1$ modes.}. This memory contribution can be computed via the most important oscillatory high-frequency waveform modes, noted $h_{\ell m}^H$, used in this work as Eq. (39) in~\cite{Zosso:2025memory}:

\begin{eqnarray}
  h_{20}^{\rm (mem)}(t) =\int_{-\infty}^tdt' \Bigg[\frac{1}{7}\sqrt{\frac{5}{6\pi}}|\dot h^H_{22}|^2 -\frac{1}{14}\sqrt{\frac{5}{6\pi}}|\dot h^H_{21}|^2 \nonumber\\ +\frac{5}{4\sqrt{42\pi}}\left(\dot h^H_{22}\dot h^{*H}_{32}+\dot h^{*H}_{22}\dot h^H_{32}\right) \nonumber\\
  +\frac{1}{2\sqrt{10\pi}}\left(\dot h^H_{33}\dot h^{*H}_{43}+\dot h^{*H}_{33}\dot h^H_{43}\right) \nonumber\\ -\frac{2}{11}\sqrt{\frac{2}{15\pi}}|\dot h^H_{44}|^2\Bigg]\,.
  \label{eq:BMS_BalanceLaw}
\end{eqnarray}

We want to stress here that the selection of the $(2,0)$ from Eq.~\eqref{eq:memoryequation} crucially depends on the physical input of understanding gravitational memory as an inherently low-frequency contribution, implemented through the averaging $\langle ... \rangle$. Once this selection has been performed, however, the averaging can be effectively dropped.

Concerning gravitational memory, there are two distinct types of waveforms: $a.$ Historically, numerical simulations of binary black holes based on an extraction scheme of extrapolation did not resolve the memory contribution, even though such waveforms are able to accurately capture subleading higher-order GW modes. This apparent paradox can again be understood through the low-frequency nature of gravitational memory~\cite{Zosso:2025memory}. Waveforms based on such numerical relativity simulations, therefore, naturally represent a good model of memoryless signal, while the associated memory signal can efficiently be computed via Eq.~\eqref{eq:BMS_BalanceLaw} later on. $b.$ On the other hand, it is recently possible to run numerical relativity simulations based on an alternative extraction scheme known as Cauchy Characteristic Extraction (CCE) that directly computes gravitational memory. While waveforms based on such simulations naturally represent a model for the hypothesis of the presence of gravitational memory, their no-memory hypothesis counterpart is not trivially defined. In particular it is not enough to simply select out the $(2,0)$ mode of such waveforms since, in general, even for quasi-circular binary mergers, the $(2,0)$ mode also includes an oscillatory high-frequency contribution associated to the ringdown. A practical solution is, however, given by disregarding the $(2,0)$ mode of such waveforms and redefining the corresponding memory model via Eq.~\eqref{eq:BMS_BalanceLaw}. In this work, we will precisely do so.

\subsection{Waveforms and the LISA response}
\label{sec:MemLISA_waveformResponse}

\begin{figure*}
    \centering
    \includegraphics[width=0.7\linewidth]{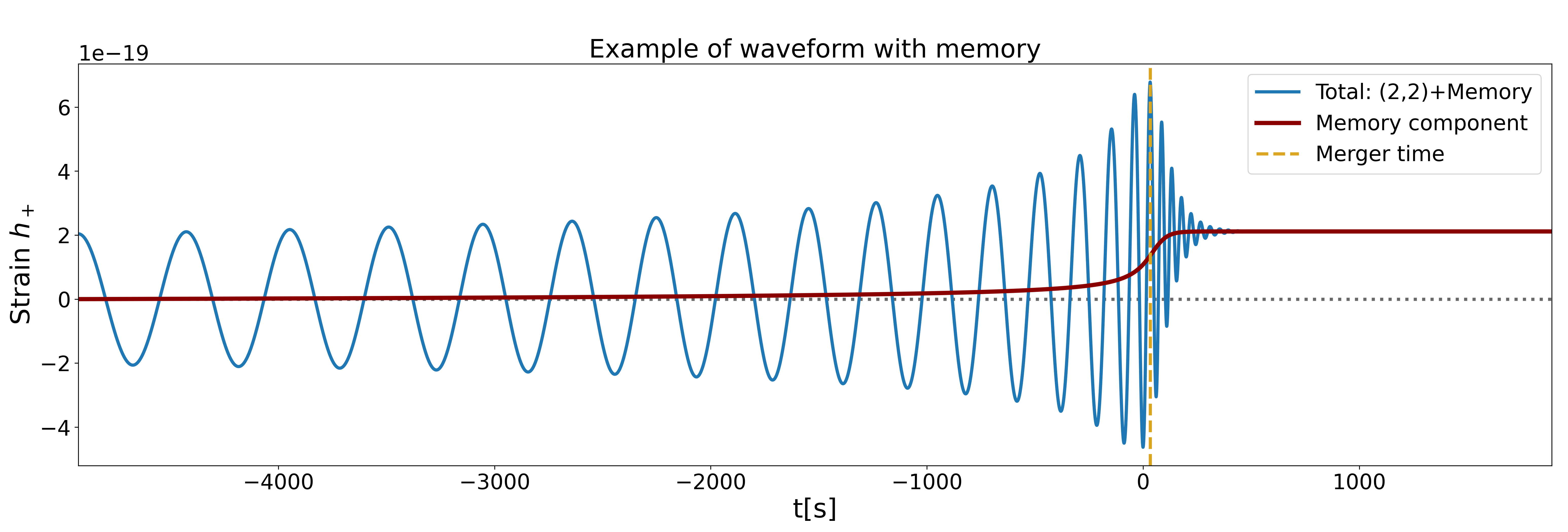}
    \caption{Example of the + polarization of a time-domain waveform with memory effect using the waveform {\tt NRHybSur3dq8\_CCE}. The blue curve shows the total signal ($(2,2)$-oscillatory + memory) and its associated $(2,0)$ memory component in red. 
    The parameters are $Q=1.5$, $\chi_{\mathrm{1z}}=0.7$, $\chi_{\mathrm{2z}}=0.7$, $M=10^6 ~\mathrm{M_\odot}$, $d_\mathrm{L}=10^4$ Mpc, $\iota=\frac{\pi}{2}$, $\varphi_{\mathrm{ref}}=0$, $\psi = 0$.}
    \label{fig:WaveformWithMem}
\end{figure*}

For comparison purposes, two waveforms models are used. Both are time-domain models originally generated in their spin-weighted spherical harmonic basis -- needed for memory computation -- that we convert into the polarization basis, which is the physical input seen by LISA. When introducing a given mode $(\ell,m)$, we also include $(\ell,-m)$, taking advantage of the symmetry of non-precessing binary systems given by $h_{\ell, -m} = (-1)^\ell h_{\ell, m}^{\ast}$.

The first waveform model we consider is \texttt{NRHybSur3dq8\_CCE}~\cite{NRHybSur3dq8_CCE}, a hybridized surrogate model from the publicly available python package \texttt{GWSurrogate}~\cite{GWSurrogate}. This surrogate model is built using numerical relativity (NR) waveforms that are mapped to the post-Newtonian (PN) BMS frame and then hybridized with effective-one-body phase-corrected PN waveforms, making it just as accurate as the surrogate built using the extrapolation scheme, \texttt{NRHybSur3dq8}~\cite{NRHybSur3dq8}, but with the advantage of exhibiting the memory effect.
This waveform contains several HMs, including the $(2,0)$ mode with memory (type $b.$). While this waveform is great for its accuracy compared to NR, it is rather slow compared to other waveform models, especially when HMs are included. This limitation might be problematic when performing a Bayesian analysis, given the large number of waveform evaluations required.
As a consequence, we will use only the dominant component of the \texttt{NRHybSur3dq8\_CCE} waveform, namely the $(2,2)$ mode, and, as explained above, compute the associated memory effect contribution separately through Eq.~\eqref{eq:BMS_BalanceLaw} by only keeping the dominant first term in the formula.

The second waveform model we consider is \texttt{SEOBNRv5HM}~\cite{SEOBNRv5HM} from the publicly available \texttt{PySEOBNR}~\cite{pySEOBNR} package. This waveform model is built using the effective-one-body formalism and calibrated to NR.
We use this model to cross-check the results obtained with \texttt{NRHybSur3dq8\_CCE} and take several HMs into account, namely the $(2,1)$, $(3,3)$, $(3,2)$, $(4,4)$, and $(4,3)$ modes. Despite the inclusion of HMs, \texttt{SEOBNRv5HM} does not include the inherent memory effect (type $a.$) and will undergo the same memory computation process as \texttt{NRHybSur3dq8\_CCE}.

After the computation of the memory through Eq.~\eqref{eq:BMS_BalanceLaw} for both waveforms, with all the modes available in a given waveform, we transform to the polarization basis via Eq.\eqref{eq:ModesToPolar}. And finally, we rotate this source-frame polarization basis to the detector-frame basis by fixing the freedom in the polarization angle $\psi$, whose rotation on the waveform is given by
\begin{equation}
    \left\{\begin{matrix}
    h_+\rightarrow h_+ \cos(2\psi)-h_\times \sin(2\psi) \\
    h_\times \rightarrow h_\times \cos(2\psi)+h_+ \sin(2\psi).
    \end{matrix}\right.
    \label{eq:PolarizationAngle}
\end{equation}
This is what we physically observe when gravitational radiation reaches our detector.

In Table~\ref{tab:parameters_definition}, we introduce the physical parameters needed to describe the MBHB for the generation of waveforms and, therefore, the subsequent associated response of LISA. We also provide the range of the parameters, which are chosen based on either physical considerations (e.g., inclination) or limitations of the waveforms used (e.g., mass ratio). There are no limitations for $d_\mathrm{L}$; the indicated range is simply the one used for the analysis. We use the conventions defined in the LISA Rosetta Stone Convention document of the LISA Distributed Data Processing Center (DDPC)~\cite{LISA_RosettaStone}.

\begin{table*}
    \begin{tabular}{c|c|c}
        Parameter & Definition & Considered range \\
        \hline
        \multirow{2}{*}{Total mass $M$} & Total mass of the binary, & \multirow{2}{*}{$M \in [10^4, 10^8] ~\mathrm{M_\odot}$ } \\[-0.8ex]
                                     & seen in the detector frame. & \\[0.6ex]
        \multirow{2}{*}{Mass ratio $Q$} & $Q = \frac{m_1}{m_2}$, & \multirow{2}{*}{$Q \in [1, 8]$} \\
                                 & where $m_1 > m_2$, and $m$ the mass of a given BH &  \\[0.6ex]
        \multirow{2}{*}{Spin $\chi_{\mathrm{(1/2)z}}$} & Dimensionless spin component of BH 1/2, &\multirow{2}{*}{$\chi_\mathrm{z} \in [-0.8, 0.8]$} \\[-0.8ex]
                                 & along the orbital angular momentum &  \\[0.6ex]
        Luminosity distance $d_\mathrm{L}$ & Luminosity distance to the binary & $d_\mathrm{L} \in [10^2, 10^6]$ Mpc \\[0.6ex]
        \multirow{2}{*}{Inclination $\iota$} & Apparent inclination of binary ecliptic plan & \multirow{2}{*}{$\iota \in [0,\pi]$} \\[-0.8ex]
                                 & ($0=$ face-on, $\frac{\pi}{2}=$ edge-on) &  \\[0.6ex]
        Polarization angle $\psi$ & Transforms source $(h_+, h_\times)$ to detector $(h^\prime_+, h^\prime_\times)$ & $\psi \in [0,\pi]$ \\[0.6ex]
        Right ascension $\alpha$ & Equatorial sky coordinate & $\alpha \in [0, 2 \pi]$ \\[0.6ex]
        Declination $\delta$ & Equatorial sky coordinate & $\delta \in [-\frac{\pi}{2}, \frac{\pi}{2}]$ \\[0.6ex]
        \hline
        Secondary parameter & Definition & --- \\
        \hline
        Redshift $z$ & Computed with $d_\mathrm{L}$ using Planck2015 model~\cite{Planck2015, astropy:2013, astropy:2018, astropy:2022} & --- \\
        Source-frame total mass & $M_{\mathrm{source}} = \frac{M}{1+z}$ & --- \\
        Latitude $\beta$ & Ecliptic sky coordinate (legacy coordinate) & $\beta \in [-\frac{\pi}{2},\frac{\pi}{2}]$ \\[0.6ex]
        Longitude $\lambda$ & Ecliptic sky coordinate (legacy coordinate) & $\lambda \in [0, 2\pi]$ \\[0.6ex]
    \end{tabular}
    \caption{Definition of all the physical parameters needed to generate a waveform (and the associated response) with the range taken into account for this study.}
    \label{tab:parameters_definition}
\end{table*}

In Fig.~\ref{fig:WaveformWithMem}, we show an example of a time-domain waveform for GWs emitted by a non-precessing MBHB. 
This figure shows the oscillatory part of the $(2,2)$ mode with its associated memory effect contribution, i.e., the $(2,0)$ mode component introduced in Eq.~\eqref{eq:BMS_BalanceLaw}.
These parameters are chosen, following Ref.~\cite{Henri_Memory_Paper}, to obtain a significant memory effect, with a mass ratio $Q$ close to one and high positive spin components which, among other things, contribute to a higher amplitude for the memory effect.
The choice of an edge-on inclination $\iota = \frac{\pi}{2}$ and a polarization angle $\psi=0$ allows the memory amplitude to be maximal (depending on $\iota$), and to have only a plus-polarization for both the $(2,2)$ mode and the memory component, or $h_\times = 0$.
As can be seen in Fig.~\ref{fig:WaveformWithMem}, the memory effect primarily builds up during the late merger phase because it corresponds to the maximum energy release in GWs.

The next step is to include the response and TDI processing so we obtain L1-like data, which is how the TDI variable outputs are called. This is usually performed with the LISA software suite {\tt GWresponse}~\cite{lisa_gwresponse} and {\tt PyTDI}~\cite{PyTDI} to evaluate the response of LISA laser links and then compute the TDI variables. These computations also require orbits of LISA's spacecrafts, which are simulated using the {\tt LISAorbits} package~\cite{lisa_orbits}.

In this work, we computed the orbits once and then used the fast version of the response+TDI, which was implemented in the {\tt LISARing} software --this fast response is a prototype of the {\tt LISAX} implementation \cite{LISAX}, which is not yet public-- to save calculation time in view of the subsequent Bayesian analysis which is computationally time demanding. The detailed analysis of how LISA responds to the memory component can be found in Inchauspé et al.~\cite{Henri_Memory_Paper}. For the sake of clarity and completeness, an example of the post-processed signal shown in Fig.~\ref{fig:WaveformWithMem} can be found in Appendix~\ref{App:LisaResponse}, Fig.~\ref{fig:TimeTDIA}, along with technical notes to ensure reproducibility.
We perform a Fourier transform on the resulting time-domain TDI channels, as the frequency domain is more convenient for SNR and likelihood computation. To limit spectral leakage, we use the Planck-taper window~\cite{Planck-taper_Windowing}, which is often adopted in GW analyses.
We make use of the analytical Power Spectral Density (PSD) of the noise model~\cite{LISA_SciRD} as a reference to evaluate the noise level.

\subsection{Templates construction} 
\label{sec:MemLISA_template}

To study the detectability of the memory effect, we start by constructing simulated MBHB datasets (including memory) in which we will search for the injected memory.
These datasets, sometimes abbreviated as `data', represent mock L1 output from LISA and, therefore, are generated with a noise realization.
To evaluate our ability to identify memory, we will build two noiseless models (also called templates), one including memory and one without. Once the waveform is modulated by the response of LISA, we can search for memory in the data. To compare the two models, we make use of the Bayes factor computed through Bayesian inference or through a likelihood estimation that we introduce below in Section~\ref{sec:BayesianAnalysis}.
 
Comparing a dataset with the two templates allows us to know which template is considered a better fit to the data. In our case, a significant preference for the memory template when data contains memory allows us to claim memory detection.

To build a dataset or a template, we start by choosing a set of parameters for the MBHB and generating the waveform using one of the two waveform models with their mode decomposition.
We, then, add the memory component in the relevant case (mock data or memory template), calculated from Eq.~\eqref{eq:BMS_BalanceLaw}, and project the GW onto the detector polarization basis to obtain the incoming signal seen by LISA.

\begin{figure}[H]
    \centering
    \includegraphics[width=1\linewidth]{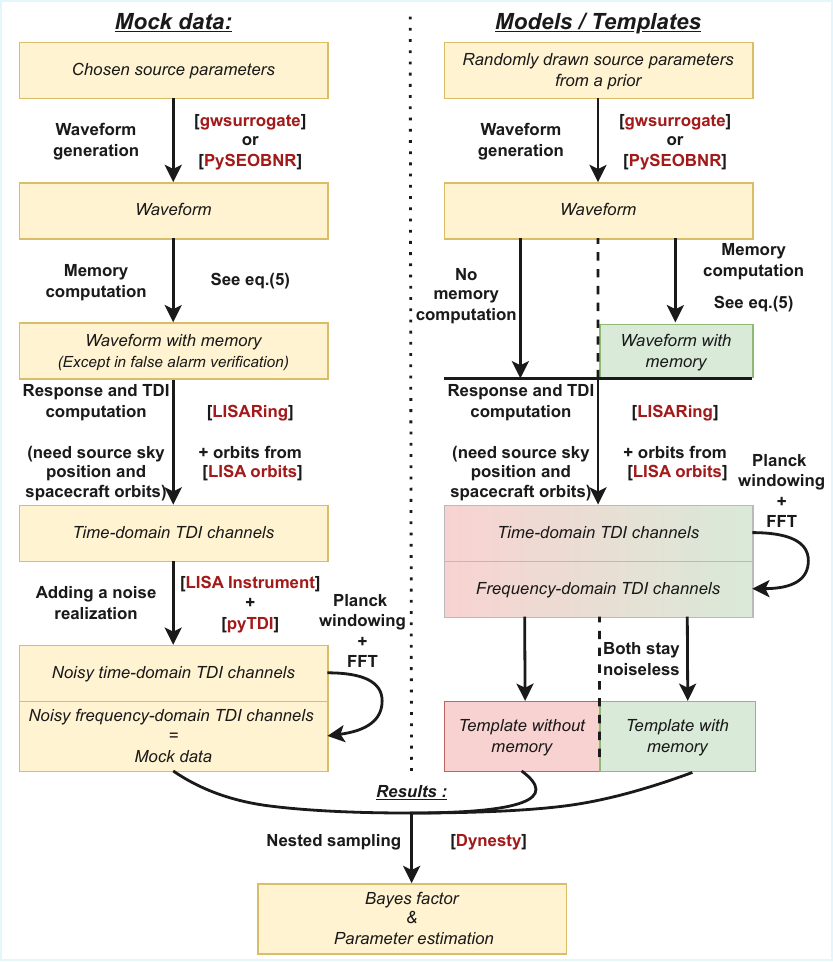}
    \caption{Summarized steps to obtain mock data and templates. Red names indicate the package use for a given process.}
    \label{fig:Generation_diagramm}
\end{figure}

Providing the orbit and the sky position, {\tt LISAring} will compute the response and the time-domain TDI channels A, E, and T.
The final step is to apply Planck windowing and perform a Fourier transform to obtain the frequency-domain TDI channels.
Fig.~\ref{fig:Generation_diagramm} summarizes the whole process.

The noise realization is obtained with the {\tt LISAInstrument} package~\cite{LISAInstrument, LISAInstrumentCompagnionPaper} of the LISA simulation suite to simulate the different types of noise affecting LISA data. We directly convert them in TDI time-domain channels with pyTDI~\cite{PyTDI}.
Here, we took into account the noise from the optical metrology system (OMS) and the acceleration noise, as these are the main sources of noise in LISA after the laser noise has been removed by TDI.
The formulae and values describing these noise sources are taken from the Science Requirements Document~\cite{LISA_SciRD}. 
In this work, we only consider the noise produced by the instrument, neglecting the significant galactic confusion noise. 
From a first estimation, using the simulated results for 4 years of observation of this confusion noise, we expect to see a slight decrease in $\SNRmem$ for certain total masses $M$, in particular in the range $\left[ \sim10^{5.8}, \sim10^{6.5} \right] ~\mathrm{M_\odot}$ (See Fig.~\ref{fig:WaterfallPlotsSurrogateWithWDNoise} in~\ref{app:Waterfalls}), without significant changes on the detectability relation. It will also slightly reduce the number of binaries seen in the population estimation~\footnote{In Ref.~\cite{Henri_Memory_Paper}, the contribution from galactic confusion noise was included in the population analysis.}.

\section{Detectability of the memory effect}
\label{sec:Detectability}

In this section, we discuss the use of the signal-to-noise ratio (SNR) and Bayes factor (BF) as metrics to assess the detectability of the memory effect.  
In Inchauspé et al.~\cite{Henri_Memory_Paper}, the SNR of the memory component was used as an indicative measure. Here, we extend that approach by performing a full Bayesian analysis. In particular, we use BFs to determine which model—memory or no-memory—best fits a noisy dataset. This provides a rigorous criterion for establishing whether, for a given set of source parameters (and a specific noise realisation), the memory contribution can be distinguished and claimed as detected.

However, carrying out a full Bayesian analysis using dynamical nested sampling is computationally expensive, making it impractical to explore the entire parameter space of our binaries.  
Our goal is, therefore, to investigate whether BFs can be used to calibrate a threshold in $\SNRmem$, enabling a fast and efficient estimate of the regions of the MBHB parameter space where memory is expected to be detectable.

\subsection{Computation of the signal-to-noise ratio}
\label{sec:Detectability_SNR}
 
We begin by computing the SNR for a given set of parameters.  
From the TDI processing we obtain three independent frequency-domain channels, and we compute the SNR $\rho_\alpha$ for each $\alpha \in \{A,E,T\}$.  
Since the AET channels are mutually independent, the total SNR, $\SNRtot$, is given by the square root of the quadratic sum of the individual $\rho_\alpha$.

The formal expression is
\begin{eqnarray}
    &\SNRtot = \sqrt{\sum_{\alpha \in \{A,E,T\}} \rho_\alpha^2}= \sqrt{\sum_{\alpha \in \{A,E,T\}} \langle d|d\rangle_\alpha} \qquad \\
    & {\qquad}= \sqrt{\sum_{\alpha \in \{A,E,T\}}
      \left( 4\,{\rm Re} \int_{f_{\rm min}}^{f_{\rm max}}
      \frac{\tilde{d}_\alpha^{*}(f)\,\tilde{d}_\alpha(f)}{S_\alpha(f)}\, df \right)^2 } \nonumber\,
    \label{eq:SNR_formula}
\end{eqnarray}
where $\tilde{d}_\alpha(f)$ is the Fourier transform of the output of channel $\alpha$, and $S_\alpha(f)$ is the corresponding analytical noise power spectral density (PSD).  
We use the standard inner product,
\begin{equation}
    \langle a|b\rangle = 4\,{\rm Re}\int_{f_{\rm min}}^{f_{\rm max}}
    \frac{\tilde{a}^{*}(f)\,\tilde{b}(f)}{S(f)}\, df ,
    \label{eq:inner_product}
\end{equation}
with $(f_{\rm min},\,f_{\rm max}) = (10^{-4},\,10^{-1})~\mathrm{Hz}$, which covers the frequency band in which LISA is most sensitive.

The total SNR, which includes both the oscillatory and memory components, depends on the duration of the inspiral phase of the MBHB under consideration.  
However, since our analysis focuses on the memory signal, which is predominantly accumulated during the merger (see Fig.~\ref{fig:WaveformWithMem}), we evaluate the memory contribution from the final $\sim 10$ cycles of the inspiral up to the end of the ringdown.
Extending the inspiral further does not significantly affect the SNR of the memory.

\begin{figure}[H]
    \centering
    \includegraphics[width=0.9\linewidth]{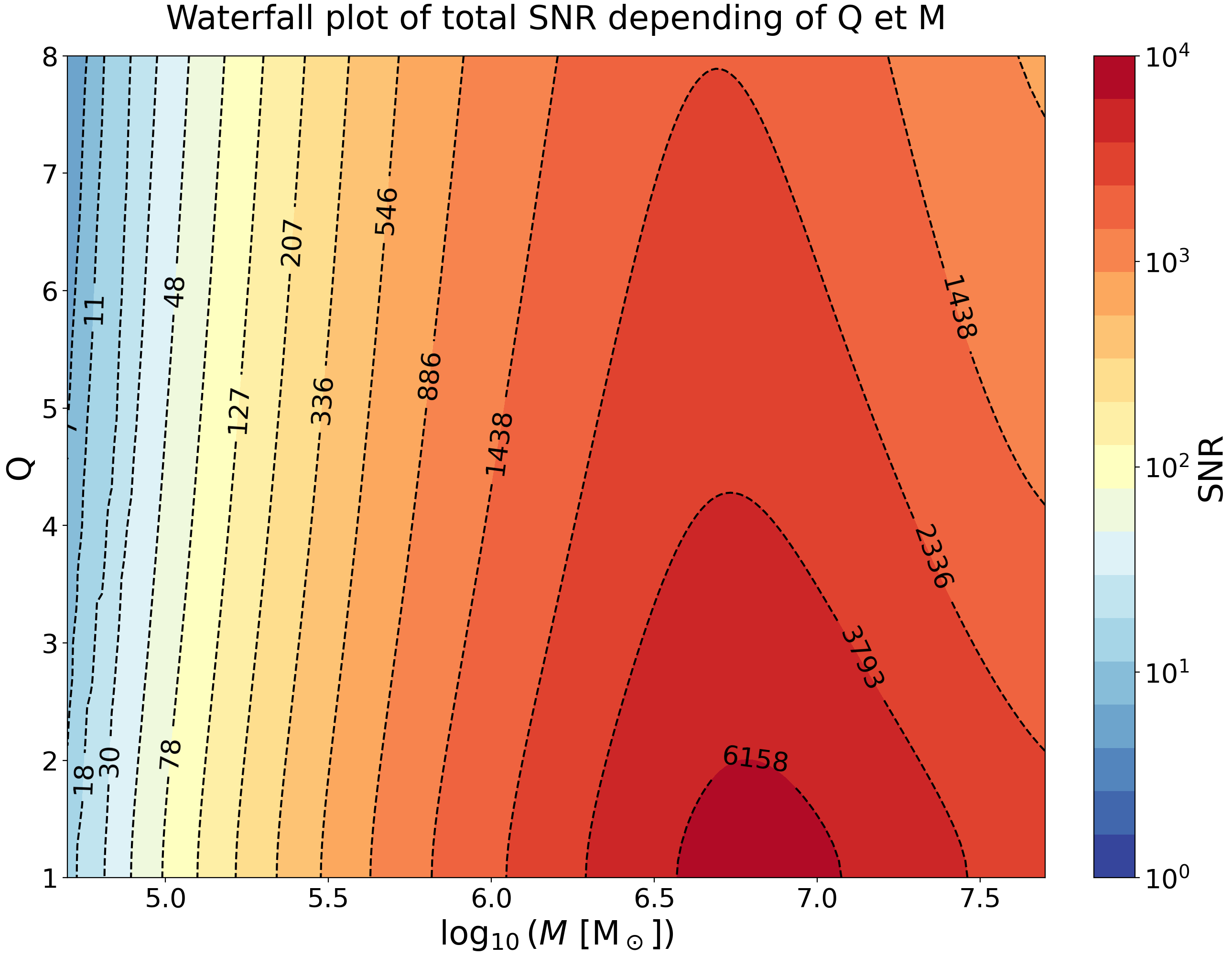}
    \includegraphics[width=0.9\linewidth]{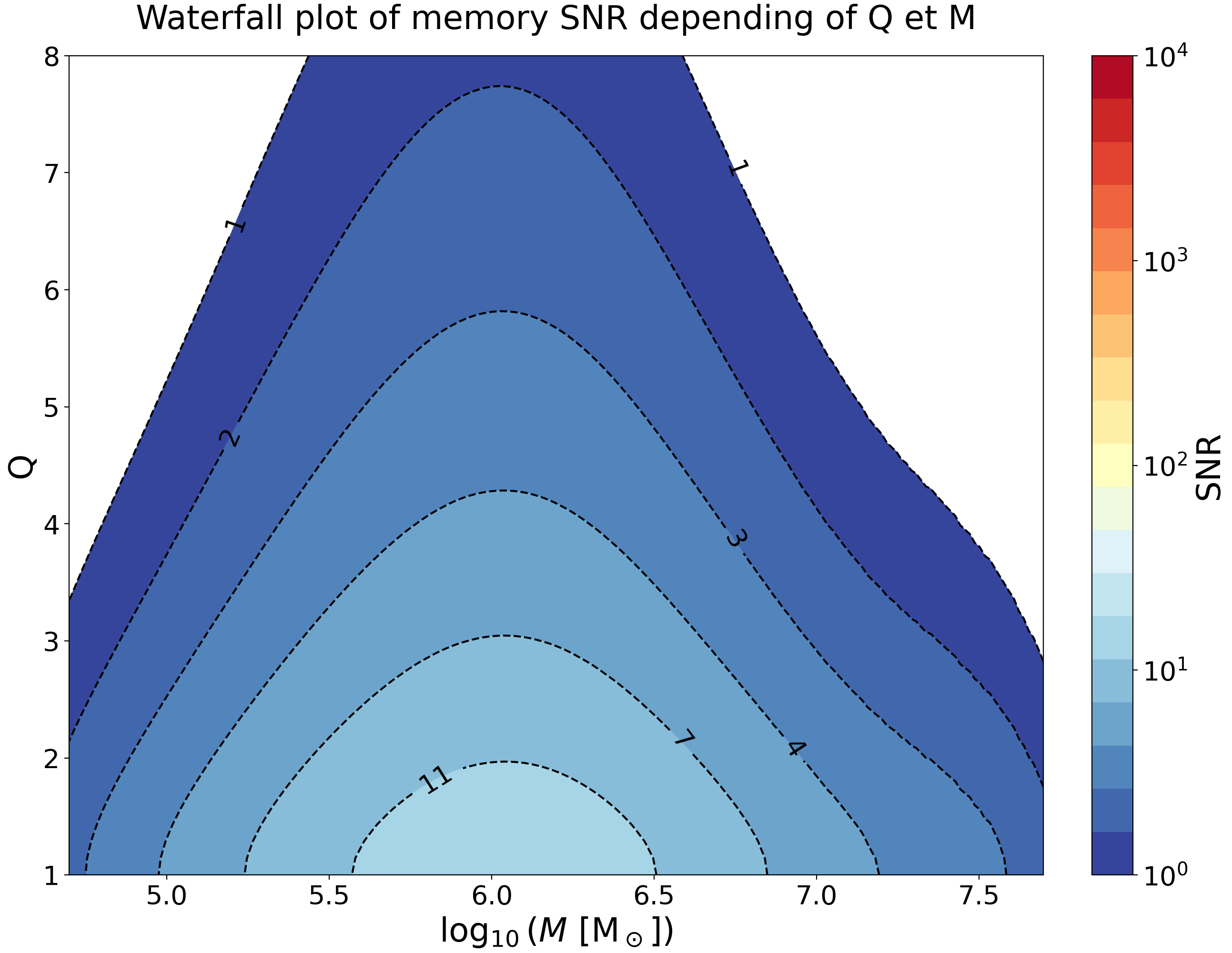}
    \caption{$\SNRtot$ (top subfigure) and $\SNRmem$ (bottom subfigure) depending on the total source mass $M$ and the mass ratio $Q$. Because of the different frequency content between the oscillatory and the memory signal, the peak of the sensitivity is different in the two cases.
    Here we used the {\tt NRHybSur3dq8\_CCE} waveform and the following parameters: $\chi_{\mathrm{1z}} = \chi_{\mathrm{2z}}=0.4$, $\iota = \pi/3$, $d_{\mathrm{L}} = 10^4 ~\mathrm{Mpc}$, $\varphi_{\mathrm{ref}} = 1$, $\psi = 0$, $\alpha = 0.74$, $\delta = 0.29$.}
    \label{fig:WaterfallPlotsSurrogate}
\end{figure}

We denote by $\SNRmem$ the SNR associated with the memory component alone.
With this setup we compute the SNR of the full signal and of the memory component for different values of parameters, and we represent them in what are known as `waterfall' plots.
Fig.~\ref{fig:WaterfallPlotsSurrogate} shows the SNR contours using the {\tt NRHybSur3dq8\_CCE} waveform for different mass ratio and total mass parameters. 
Despite some differences in the method such as memory computation (using Eq.\eqref{eq:BMS_BalanceLaw} in this work, or using {\tt GWMemory}~\cite{GWmemory} in~\cite{Henri_Memory_Paper}), the results are in good agreement with those obtained by Inchauspé et al.~\cite{Henri_Memory_Paper}.
The parameters used in the waterfall plots have been chosen in such a way that the memory part is quite substantial, albeit without being in the most optimistic scenario. Compared to~\cite{Henri_Memory_Paper}, we use a set of parameters that lies between conservative and optimistic scenarios. In particular, the choice of the sky position corresponds to a sky position with an average $\SNRmem$ value.
In section~\ref{sec:Catalogs}, we will put this in perspective by referring to catalogues of sources.

We also construct waterfall plots using the {\tt SEOBNRv5HM} waveform where the SNR is computed with the same parameter values as for the {\tt NRHybSur3dq8\_CCE} case in Fig.~\ref{fig:WaterfallPlotsSurrogate} for the sake of consistency and comparison and for the same range in mass ratio $Q$ and total mass $M$.
The inclusion of HMs benefits the memory effect as a slight increase in SNR due to the HMs contribution to the memory through Eq.\eqref{eq:BMS_BalanceLaw}.
The results and discussion can be found in~\ref{app:Waterfalls}.

As in Inchauspé et al.~\cite{Henri_Memory_Paper}, we find a wide region of parameter space where $\SNRmem$ is large enough to expect detection.
We now turn to the Bayesian analysis to evaluate our ability to disentangle the memory signal from the oscillatory part and define a proper detectability criterion.

\subsection{Bayesian analysis and memory detectability criteria}
\label{sec:BayesianAnalysis}
\subsubsection{Bayesian analysis}
Working within the Bayesian framework offers a double benefit as it allows for both parameter estimation and model comparison.
According to the Bayes theorem, one can express the posterior probability distribution $p(\bm{\theta}|d, m)$ of the source parameters $\bm{\theta}$ given the data $d$ and the model $m$ as:
\begin{equation}
    p(\bm{\theta}|d, m) = \frac{p(d|\bm{\theta}, m) p(\bm{\theta}|m)}{p(d|m)}
    \label{eq:BayesTheorem}
\end{equation}

During LISA science operations, data $d$ will come from the L01 pipeline of LISA \footnote{In a nutshell, the L01 pipeline process the single-link phase measurements of LISA (L0 data), essentially applying relevant time-delays and data streams combination in order to produce the synthetic interferometer (TDI) output data (L1 data).}. Here the mock data are generated through the process described Fig.~\ref{fig:Generation_diagramm}, with a GW signal with source parameters $\bm{\theta}_{\rm source}$ and a noise realization.
The models used to compare the data are built with the same waveform model, but no noise realization.
Thus, in Eq.~\eqref{eq:BayesTheorem}, the $\mathcal{L} = p(d|\bm{\theta}, m)$ term corresponds to the likelihood, i.e. the probability of having this particular dataset $d$ given a certain model $m$ evaluated with the parameter set $\bm{\theta}$. 
The $p(\bm{\theta}|m)$ term is the prior distribution of the parameter space for model $m$, and the $\mathcal{Z} = p(d|m)$ corresponds to the evidence and describes how well a given model $m$ can describe the dataset $d$.
This evidence is the key element for performing model comparison.
We use the definition of the Bayes factor $\mathcal{B}$, given by the ratio between the evidence of two given models, in our case the model with memory (labelled by ${ o+{\rm mem}}$ for oscillatory plus memory components) and without memory (labelled by $o$):
\begin{equation}
    \mathcal{B}_{\mathrm{mem}} = \frac{\mathcal{Z}_{o+ {\rm mem}}}{\mathcal{Z}_{o}}
    \label{eq:BayesFactor}
\end{equation}

Therefore, we need to compute the evidence and the likelihood to obtain the posterior distribution and the Bayes factor.
In practice, we prefer to calculate the ln-likelihood\footnote{In case of omission of the base, please consider $\log$ as the base-10 logarithm. When needed, the natural logarithm is always written as $\ln$. Later on, we will prefer log to ln-likelihood.} rather than the likelihood, and we express it (up to a constant) as follows:
\begin{equation}
    \ln\mathcal{L} \propto -\frac{1}{2}  <d - m(\bm{\theta})|d - m(\bm{\theta})>
    \label{eq:Likelihood}
\end{equation}
where we used the inner product defined in Eq.\eqref{eq:inner_product}.

For a model $m$, the evidence can be computed as:
\begin{equation}
    \mathcal{Z} = p(d|m) =\int \mathcal{L}(d|\bm{\theta}, m) p(\bm{\theta}|m)d\bm{\theta} = p(m|d) p(m)
    \label{eq:Evidence}
\end{equation}

Eq.~\eqref{eq:Evidence} provides the relation between the evidence and the probability $p(m|d)$ to have a model $m$ given the observed data $d$. Here, we will not set a prior preference between the models, so the inner probability of having a model $p(m)$ will be the same for both and cancel out in the ratio.

We now turn to the practical application of the above formalism.
We use the dynamical nested sampling method~\cite{DynamicNestedSampling} to compute the evidence and estimate the posterior distribution, using {\tt Dynesty}~\cite{Dynesty, DynamicNestedSampling, Skilling_2004, Skilling_2006, Speagle_2020, Multinest}. 
Unlike Markov Chain Monte Carlo (MCMC), the nested sampling approach allows for obtaining the evidence of a model against the data as its primary outcome, as well as the posterior distribution as a by-product.
The parameter estimation from the posterior distribution will be detailed in Section~\ref{sec:ParamEstimation}, where we analyse the effect of adding memory on parameter reconstruction as well as the precision with which one can infer the characteristics of the memory strain.

The comparison between models is performed by running the {\tt Dynesty} nested sampler twice: once for the model $m$ including memory and once for the model without memory.  
After both runs have converged, {\tt Dynesty} provides an estimate of the log-evidence $\ln \mathcal{Z}$ for the corresponding model. We then compute the BF for the presence of memory as
\begin{equation}
    \log_{10} \mathcal{B}_{\rm mem}
    = \frac{\ln \mathcal{Z}_{o+ {\rm mem}} - \ln \mathcal{Z}_{\rm o}}{\ln 10} .
    \label{eq:logBF}
\end{equation}
To interpret the resulting BFs, we adopt the Jeffreys scale (see Appendix~B of~\cite{Jeffreys_1998}), summarised in Table~\ref{tab:JeffreysScale}.  
According to this scale, the BF—or its logarithm—falls into one of five categories, ranging from ``barely worth mentioning'' to ``decisive'' evidence in favour of one model.

The interpretation naturally extends to negative values of $\log_{10}\mathcal{B}_{\rm mem}$ or to values in the range $[0,1]$, since the same criteria apply in favour of the alternative model.  
For instance, a result $\log_{10}\mathcal{B}_{\rm mem} = -2.5$ indicates ``decisive'' evidence for the model without memory, corresponding to $\log_{10}\mathcal{B}_{\rm no\,mem} = 2.5$.

\begin{table}[H]
 \begin{ruledtabular}
    \begin{tabular}{ccc}
        $\mathcal{B}$ & $\log_{10} \mathcal{B}$ & Interpretation \\
        \hline
        $[1, 3]$ & $[0, \frac{1}{2}]$ & Barely worth mentioning \\
        $[3, 10]$ & $[\frac{1}{2}, 1]$ & Substantial \\
        $[10, 32]$ & $[1, \frac{3}{2}]$ & Strong \\
        $[32, 100]$ & $[\frac{3}{2}, 2]$ & Very strong \\
        $[100, +\infty[$ & $[2, +\infty[$ & Decisive \\
    \end{tabular}
    \caption{Jeffreys scale to interpret the (log) Bayes factor result. }
    \label{tab:JeffreysScale}
 \end{ruledtabular}
\end{table}

We consider a detection of the memory when we reach a decisive evidence according to the Jeffreys scale, i.e. $\log_{10}\mathcal{B}_{\mathrm{mem}} > 2$. We could then repeat the same procedure for various $\bm{\theta}$ and build maps similar to the $\SNRmem$ waterfall plot (see Fig.~\ref{fig:WaterfallPlotsSurrogate}) but with BF values, assessing detectability regions.
\begin{figure}[H]
    \centering
    \includegraphics[width=0.95\linewidth]{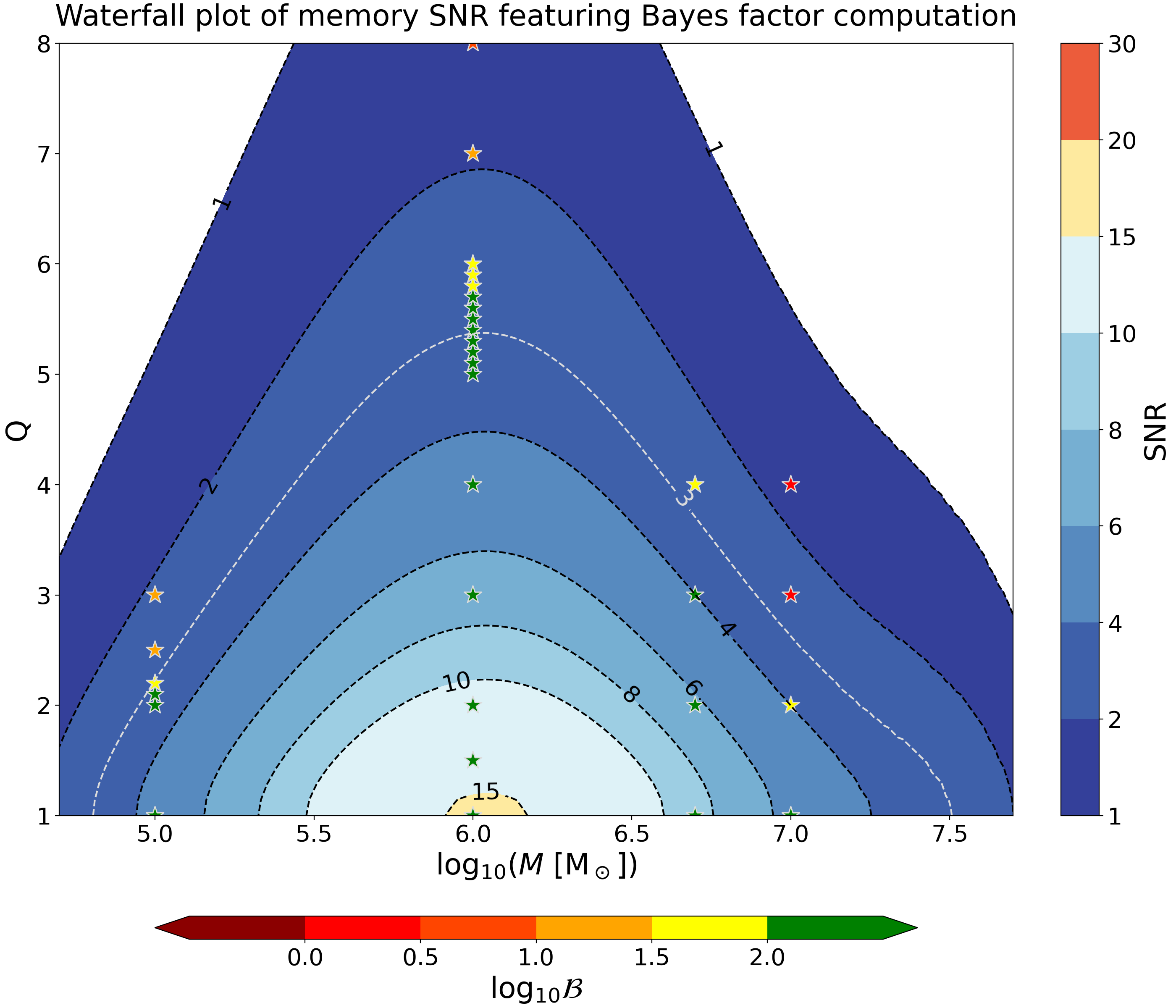}
    \caption{Memory waterfall plot from the Fig.~\ref{fig:WaterfallPlotsSurrogate} with stars corresponding to the computed $\log_{10}\mathcal{B}$. 
    The colour of the stars corresponds to the Jeffreys scale (Table~\ref{tab:JeffreysScale}) as indicated by the colour-bar under the figure. 
    The light gray dashed line represents the ISO-SNR contour $\SNRmem = 3$.
    This plot used the {\tt NRHybSur3dq8\_CCE} waveform and the same parameters as Fig.~\ref{fig:WaterfallPlotsSurrogate}}
    \label{fig:MemoryWaterfallPlotWithBF}
\end{figure}
However, as previously mentioned, the time required to compute the evidence is prohibitively long to allow for complete mapping~\footnote{All the runs were done on IRFU's cluster. For each point, we need two runs, one for each model. A run with a 3000 live points sampler and 100 CPUs used for parallel computing typically takes between 10 and 30 hours.}. 
To overcome this issue, we restrict ourselves to computing fewer points and superimpose them on the relevant SNR waterfall plot as shown in Fig.~\ref{fig:MemoryWaterfallPlotWithBF}.
The stars in Fig.~\ref{fig:MemoryWaterfallPlotWithBF} indicate the points in the ($Q$, $M$) plane where we have performed the full BF computation.

The green stars correspond to signals with detectable memory effect.
The light gray dashed line ($\SNRmem = 3$) shows the tipping point between the favourable and unfavourable regions to detect memory, as described later in subsection~\ref{sec:VariabilityOfBF}. 
For practical reasons, each iso-mass series had a unique noise realisation per mass. This introduces a bias in the threshold estimation that will be discussed and fixed in the subsection~\ref{sec:VariabilityOfBF}.

\subsubsection{Memory detectability criteria}
\label{sec:SNRThreshold}

Having established the detectability of the memory effect through the BF computation, we explore its relationship with the SNR, aiming to define an SNR threshold and generalize the notion of detectability criteria.
To do so, we explore the variation of the BF with respect to several changes in parameter values. We start from a reference set of parameters [$M=10^6 ~\mathrm{M_\odot}$, $Q=1.5$, $\chi_{\mathrm{1z}} = \chi_{\mathrm{2z}}=0.4$, $\iota = \pi/3$, $d_{\mathrm{L}} = 10^4$ Mpc, $\varphi_{\mathrm{ref}} = 1$, $\psi = 0$, $\alpha = 0.74$, $\delta = 0.29$], and let some of them vary. We ran these BF computations with two different approaches. First, we randomly vary one parameter, such as the spin or the inclination. Second, we vary a given parameter over a wide range, as we did for the luminosity distance $d_{\mathrm{L}} \in [10^4, 10^6]$ Mpc and the mass ratio $Q \in [1,8]$. For the $Q$ series, we also calculate the BF for different total masses $M$.

\begin{figure}[H]
    \centering
    \includegraphics[width=1\linewidth]{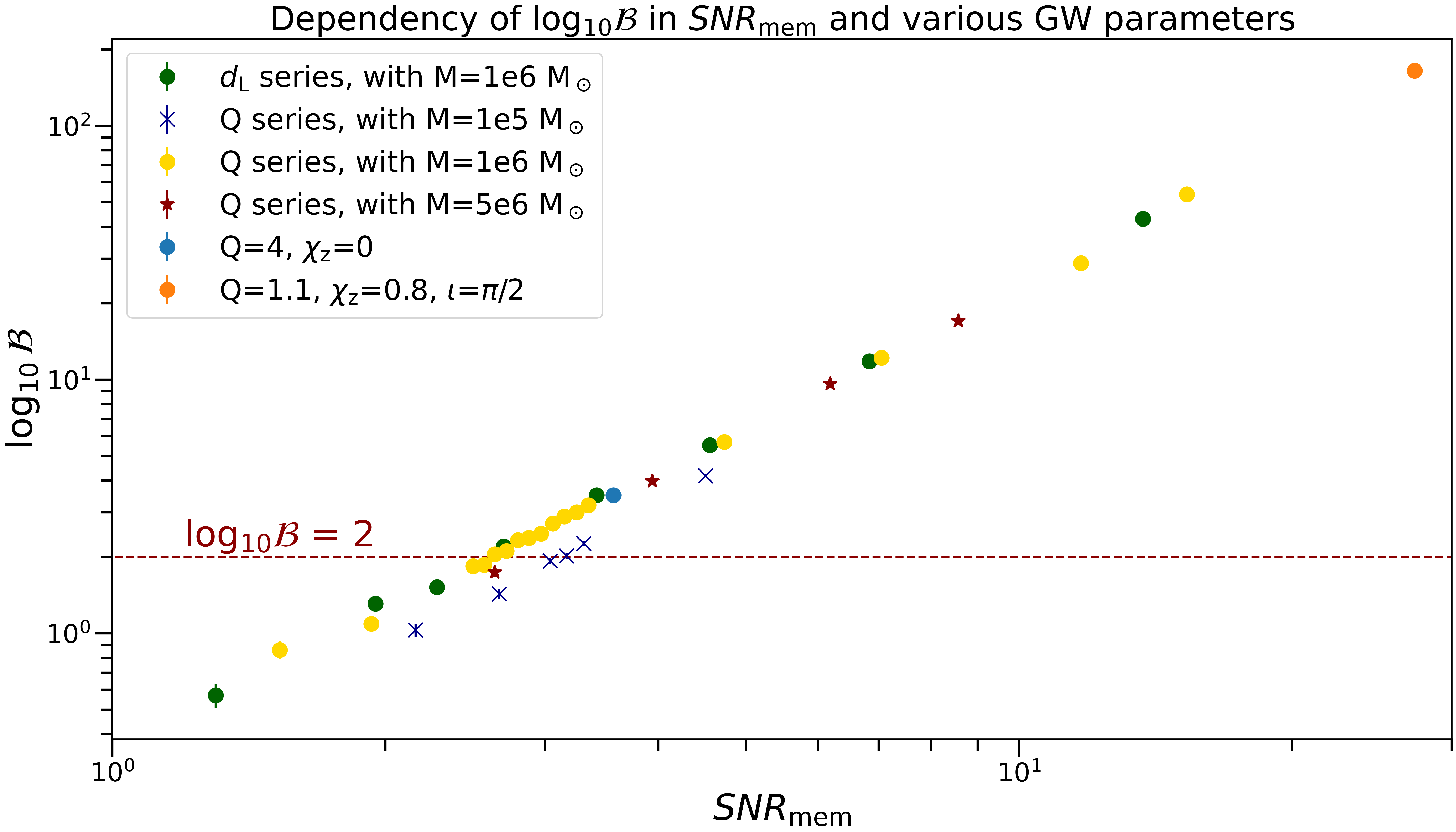}
    \caption{$\log_{10}$Bayes factor computation for different parameters. The colors stand for different parameters, except mass, which are indicated with different markers. This plot made use of {\tt NRHybSur3dq8\_CCE} waveform.}
    \label{fig:logB_dep_in_SNRmem}
\end{figure}

As a result, we obtain a first glimpse of the dependencies between $\log_{10} \mathcal{B}$ and MBHB parameters. We obtain a clear link between $\log_{10} \mathcal{B}$ and $\SNRmem$, assessing our ability to use the $\SNRmem$ independently of the effect of the oscillatory component. This is shown in Fig.~\ref{fig:logB_dep_in_SNRmem}.

In this figure, one can notice a small dependency in $M$ by comparing the points with different markers. Indeed, there is a small dispersion in the slope dependency, which differs slightly for round markers ($M=\mathrm{10^6~M_\odot}$), cross markers ($M=10^5 ~\mathrm{M_\odot}$) and star markers ($M=5\times10^6 ~\mathrm{M_\odot}$). However, we used a unique noise realization per $M$, and thus, we believe it to be the cause of the (small) dispersion.
To verify this hypothesis, additional computations of the Bayes factor were performed on noiseless mock data. The Bayes factors obtained this way do not present the bias of using one noise realization per mass and result in a unique slope independently of $M$.
For the sake of clarity, we only put two of the many ``single variation" points to limit overloading the plot\footnote{We also did some runs (like the one with $Q=4, \chi_{\mathrm{1z}} = \chi_{\mathrm{2z}} = 0$) using {\tt Bilby}~\cite{bilby_paper} implementation of {\tt Dynesty} to check the consistency of our sampling implementation. We obtain results in good agreement.}.

To conclude, this analysis provides a first indication of how $\log_{10}\mathcal{B}$ depends on $\SNRmem$, largely independent of other parameters.  
This approximately power-law behaviour will be examined in more detail in Section~\ref{sec:VariabilityOfBF}.
We can already infer that detecting the memory effect typically requires $\SNRmem \gtrsim 3$, although this criterion is based on the associated mean value of the BF, which may fluctuate due to noise realization as we explain later.
This result is comparable to that reported by Sun et al.~\cite{TianQin_memory} for the TianQin design, who find a threshold value of $\SNRmem^\mathrm{thresh}\approx2.36$, although their BF criterion appears to differ from ours, as they use $\ln\mathcal{B}\geqslant8$, based on ~\cite{Lasky_Memory_Threshold,Thrane_Talbot_2019}. This criterion is often used in LVK-related memory works.

\subsection{Variability of the Bayes factor}
\label{sec:VariabilityOfBF}
The dispersion that we observed for different masses $M$ is a hint that the noise realization will induce a statistical fluctuation on the computed BF. To study this fluctuation, the standard approach is to compute several times the $\log_{10} \mathcal{B}$ for a given $\bm{\theta}_{\mathrm{source}}$ but with various noise realizations in the dataset, including the BF computation with noiseless data. We can then reiterate for other parameter sets.
However, extensive evaluation of BF across the full parameter space and noise realizations is typically impractical for reasonable computing time and resources. 
As an alternative, one can introduce the following approximation: one can remark that the high SNR binaries that LISA is expected to detect typically exhibits a highly peaked Gaussian likelihood function around the true parameters $\bm{\theta}_{\mathrm{source}}$. 
Moreover, the prior $p(\bm{\theta}|m)$ in the calculation of the evidence in Eq.\eqref{eq:Evidence} is, in practice, the same for both models $m$.
Consequently, the prior term only acts as a constant that will cancel out in the model comparison. Then, using the peaked likelihood approximation, we can rewrite Eq.\eqref{eq:Evidence} as:
\begin{equation}
    \mathcal{Z} \propto \int \mathcal{L}(d|\bm{\theta}, m)d\bm{\theta} \approx \mathcal{L}(d|\bm{\theta}_{\mathrm{source}}, m) \Delta\bm{\theta}
    \label{eq:PeakedEvidenceApprox}
\end{equation}
Therefore,
\begin{eqnarray}
    \log_{10}\mathcal{B} = \log_{10}\mathcal{Z}_{o+\mathrm{mem}} - \log_{10}\mathcal{Z}_{o} ~~~~~~~~~~~~~~~~~~~~~~~~~\\ 
    \approx \log_{10}\mathcal{L}(d|\bm{\theta}_{\mathrm{source}}, \{o+\mathrm{mem}\}) - \log_{10}\mathcal{L}(d|\bm{\theta}_{\mathrm{source}}, \{o\}) \nonumber
    \label{eq:PeakedBayesFactorApprox}
\end{eqnarray}
This approximation states that the result of the integral is mainly driven by the value of the log-likelihood at the injection parameters, which allows to compare many noise realizations at a faster pace. 
To support this hypothesis, we compute the log-likelihood at the source parameters for the two models, with and without memory, using the parameters on which we previously computed Bayes factors, in subsection~\ref{sec:BayesianAnalysis}.
We then compare the difference between the log-likelihood of the two models and between the log-BF for the same noise realization. Note that neglecting higher-order terms is justified when focusing on the BF calculation. In particular, the second-order term in the expansion of Eq.~\eqref{eq:PeakedEvidenceApprox} is related to the Fisher matrix, which encodes information about correlations among the parameters.

\begin{figure}
    \centering
    \includegraphics[width=0.9\linewidth]{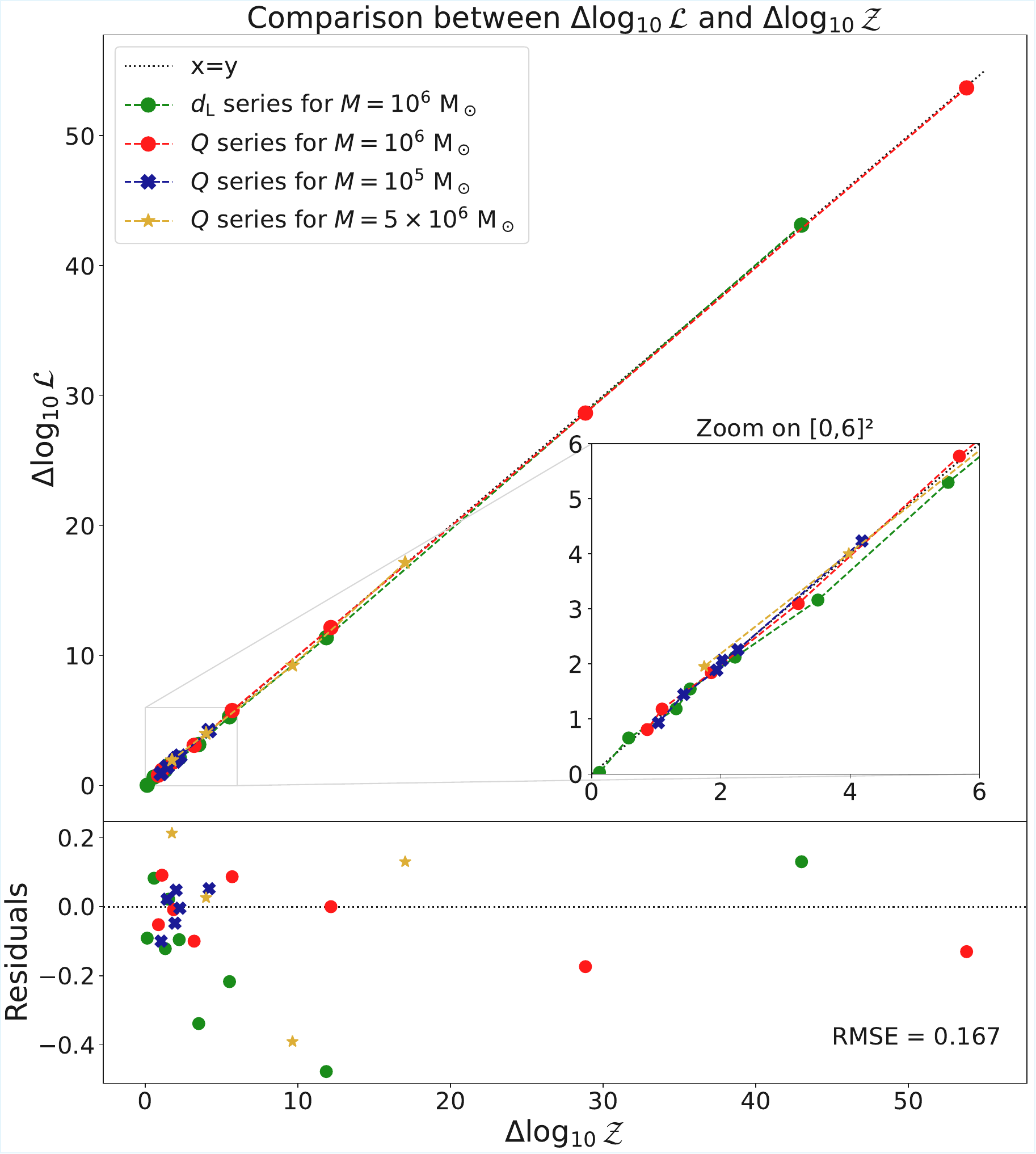}
    \caption{Comparison of the log-likelihood difference for the injected parameters $\bm{\theta}_{\mathrm{source}}$ and the log-Bayes factor for the same points as the ones computed for Fig.~\ref{fig:logB_dep_in_SNRmem}. An inset focusing on the $[0,6]^2$ region is added to distinguish the points. The lower panel shows the residuals $R = \Delta \log_{10} \mathcal{L} -  \Delta \log_{10} \mathcal{Z}$ with its associated RMSE.}
    \label{fig:LogZvsLogL}
\end{figure}

We used the $Q$ and $d_{\mathrm{L}}$ series shown in Fig.~\ref{fig:logB_dep_in_SNRmem} and compute $\Delta \log_{10} \mathcal{L}(\bm{\theta}_{\mathrm{source}})$. We plot the results against the previously computed $ \Delta \log_{10} \mathcal{Z} = \log_{10} \mathcal{B}$ in Fig.~\ref{fig:LogZvsLogL}, showing an excellent linear correlation between them, confirmed with a small value of the root mean square error (RMSE) for the residuals $R$. This supports the approximation of the strongly peaked likelihood at the true source parameters that will allow us to obtain quick estimation of $\log_{10}\mathcal{B}$.\footnote{One can also use {\tt Dynesty} trace plots to monitor the growth of the log-likelihood. In our cases, $\log_{10}\mathcal{L}$ still increase steeply when the parameter space is already highly constrained. Which further supports the proposal to use the peaked likelihood as an approximation.}

To ensure the validity of the peaked likelihood approximation, it's necessary to verify that the total SNR is always high enough. We verified the approximation down to $\SNRtot \sim 100$. One can see from Fig.~\ref{fig:WaterfallPlotsSurrogate} that this criterion is verified where we expect to have significant memory. For lower SNR, the approximation starts to break down, and we expect more variability depending also on the noise realization.

Since the computation of likelihoods is very fast (compared to the full evidence computation), we used the former introduced approximation to compute the approximated BFs of several points in the parameter space with different noise realizations\footnote{Around $\sim$ 5000 to $\sim$ 15000 realizations per set of parameters, depending on the case.}.
We obtain distributions of the $\Delta \log_{10}\mathcal{L}$ in different scenarios.
The distribution can be fitted with a simple Gaussian function, characterized by a mean value and a standard deviation $\sigma$. The mean value of $\log_{10}\mathcal{L}$ is linked to $\log_{10}\mathcal{B}$ and thus, differentiating memory effect presence, while the standard deviation represents the dispersion due to noise.
The agreement between the expected value of $\log_{10}\mathcal{B}$ for noiseless data and the mean of the distribution suggests that the noise will not lead to a systematic bias.

\begin{figure}[H]
    \centering
    \includegraphics[width=1\linewidth]{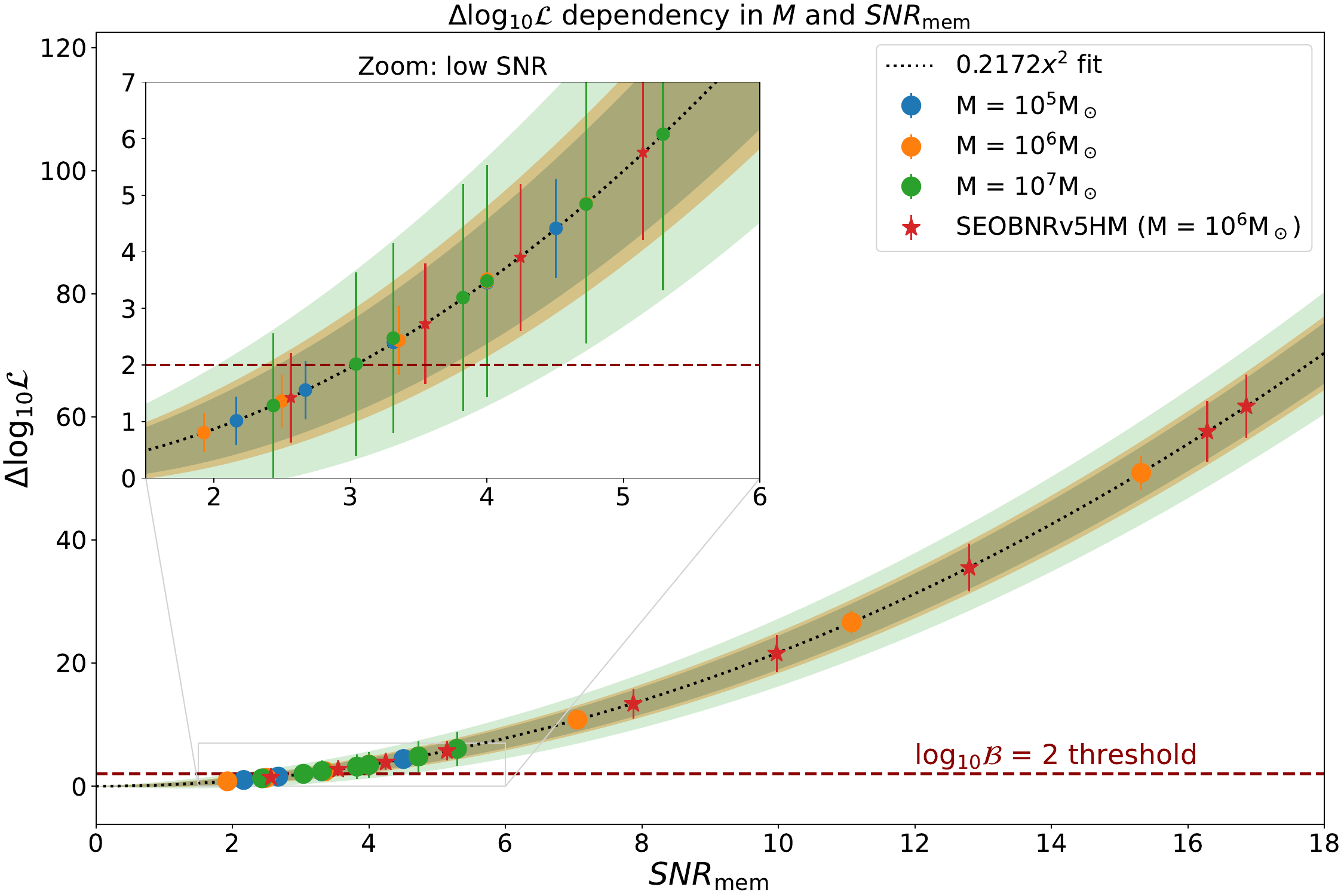}
    \caption{Mean and dispersion values of $\Delta \log_{10}\mathcal{L}(\bm{\theta}_{\mathrm{source}})$ using various parameters sets. The black dotted line shows the fitted power law linking $\SNRmem$ and $\Delta \log_{10}\mathcal{L} \approx \log_{10}\mathcal{B}$, and the red dashed line correspond to the $\log_{10}\mathcal{B} = 2$ threshold. Different total masses $M$ parameter are separated with different colours (blue for $M=10^5 ~\mathrm{M_\odot}$, orange for $M=10^6 ~\mathrm{M_\odot}$, and green for $M=10^7 ~\mathrm{M_\odot}$), both for computed points and the estimated dispersion, represented as coloured areas. 
    The dot points are computed using {\tt NRHybSur3dq8\_CCE} and are used to perform the fit. The red star points correspond to additional runs computed using {\tt SEOBNRv5HM} waveform, which includes higher modes, and are used to test the model.
    }
    \label{fig:MeanAndSigmaValues}
\end{figure}

Firstly, we can describe how the mean value of $\Delta \log_{10}\mathcal{L}$ --and thus the mean value of $\log_{10} \mathcal{B}$-- evolve with respect to a variation in the parameters and in the $\SNRmem$. We obtain that a power-law depending on $\SNRmem$ is enough to predict the mean value of the log-Bayes factor, as expected from the results of Section~\ref{sec:BayesianAnalysis}.
This can be seen in Fig.~\ref{fig:MeanAndSigmaValues}, where every point represent the mean value of set of noise realization and the error bar the associated dispersion due to noise $\sigma$. The black dotted curve shows the fitted power-law from the mean values.

Let us now focus on the dispersion $\sigma$. 
When computing the histograms for different masses, we noticed a dependence of dispersion $\sigma$ on the total mass $M$. This can be seen in Fig.~\ref{fig:MeanAndSigmaValues} by looking at the width of the error bars for points of different masses but similar $\SNRmem$. This can be explained by the fact that the total mass $M$ defines the frequency range in which the oscillatory and memory parts of the signal are in the frequency domain. As the noise PSD varies across the frequency range, the recovered signal will be more or less affected by the noise dependency on its frequency range. In particular, for memory, a higher mass like $M=10^7 ~\mathrm{M_\odot}$ can lower the signal near the limit of the sensitivity band, so a part of it is lost (i.e $f_{\mathrm{min}}=10^{-4}$Hz) and the remaining part is in a more noisy and less resolved frequency region. 

The coloured areas in Fig.~\ref{fig:MeanAndSigmaValues} correspond to the expected dispersion, slightly overestimated to be conservative on analysis choice, for the associated mass (example: orange points and orange area for $M = 10^6 ~\mathrm{M}_\odot$).

Finally, we have also added the points computed using the {\tt SEOBNRv5HM} waveform for consistency and to check if the HMs affect the power law behaviour linking $\SNRmem$ and $\Delta \log_{10}\mathcal{L} \approx \log_{10}\mathcal{B}$.
One can then see in Fig.~\ref{fig:MeanAndSigmaValues} that the points from the {\tt SEOBNRv5HM} waveform are in good agreement with the points from the {\tt NRHybSur3dq8\_CCE} waveform, i.e. with the (2,2)-mode only results.
Therefore, the presence of HMs will not affect the sampler's ability to distinguish the memory component, only the amplitude of the memory.

\subsection{Results and discussion}

The Bayes factors $\mathcal{B}$ obtained through the Bayesian analysis informed us on the MBHB parameters where we can expect to identify memory. \textit{We saw that the mean value of $\mathcal{B}$ for a set of parameters will only depend on the SNR of the memory component, considered independently of the SNR of the total signal.} Therefore we are able to define a detection threshold in $\SNRmem$, where we are favourable to identify the memory. From the usual BF threshold $\log_{10}\mathcal{B}_{\textrm{thresh}} = 2$, we can use the power law relation between $\log_{10}\mathcal{B}$ and $\SNRmem$, shown Fig.~\ref{fig:MeanAndSigmaValues}, and obtain:

\begin{eqnarray}
    \log_{10}\mathcal{B}_{\textrm{thresh}} = 2 = 0.2172 \times ({\SNRmem^{\textrm{thresh}}})^2 \nonumber\\
    \Rightarrow \SNRmem^{\textrm{thresh}} = \sqrt{\frac{2}{0.2172}}=3.034 \approx 3
    \label{eq:SNRthreshold}
\end{eqnarray}

We established the tipping point at which the memory burst should be loud enough to be favourable to detection. In addition to this result, we can use the results on the dispersion, which depends on $\SNRmem$ and total mass $M$, to evaluate the probability of detection in the region of the parameters space where noise will have a consequent impact on the detectability.
We can express this in terms of $\SNRmem$ thresholds, showing that for a higher mass we will need a higher $\SNRmem$ to be in the confident region where we expect to see memory in every event.

\begin{table}[H]
 \begin{ruledtabular}
    \begin{tabular}{cccc}
        \multirow{2}{*}{Mass} & \multirow{2}{*}{$\sigma$} & No detection & Always detected \\
         & & under $\SNRmem$ & over $\SNRmem$ \\
        \hline
        \multirow{2}{*}{$10^5~\mathrm{M}_\odot$} & $1\sigma$ & 2.47 & 3.73 \\
                                       & $2\sigma$ & 2.02 & 4.55 \\
        \hline
        \multirow{2}{*}{$10^6~\mathrm{M}_\odot$} & $1\sigma$ & 2.36 & 3.89 \\
                                       & $2\sigma$ & 1.87 & 4.93 \\
        \hline
        \multirow{2}{*}{$10^7~\mathrm{M}_\odot$} & $1\sigma$ & 2.02 & 4.55 \\
                                       & $2\sigma$ & 1.42 & 6.48 \\
    \end{tabular}
    \caption{Table of $\SNRmem$ thresholds for different masses and precision criterion at $1\sigma$ or $2\sigma$.}
    \label{tab:SNRThreshold}
 \end{ruledtabular}
\end{table}

The Table~\ref{tab:SNRThreshold} displays these results depending on the mass $M \in \{10^5, 10^6, 10^7\}~\mathrm{M}_\odot$ and the precision criterion on dispersion at $1\sigma$ or $2\sigma$.
No detection under a given $\SNRmem$ at $1 \sigma$ ($2 \sigma$) means that in a case where the $\SNRmem$ of the source is below this value, there is less than $\sim16\%$ ($\sim2.5\%$) of chance that the log-Bayes factor will reach the threshold value of 2. On the other hand, the label ``always detected over a given $\SNRmem$" means that the chance that the noise hides the memory signal for an $\SNRmem$ higher than this value is less than $\sim16\%$ ($\sim2.5\%$). In the following discussions and, particularly in Sec.~\ref{sec:Catalogs}, we will choose $\SNRmem = 5$ as the threshold value of detection for simplicity.

By looking at the values of $\SNRmem$ shown in the different waterfall plots, we can see that there is a large area of the parameter space where we can confidently expect to detect the memory effect. In addition, there is a larger area where we may detect it depending on the noise realisation.
Additionally, as the variability due to noise may appear significant, we also performed BF computations on datasets without memory and verified that the noise does not lead to false positive detection.

\begin{figure}[H]
    \centering
    \includegraphics[width=1\linewidth]{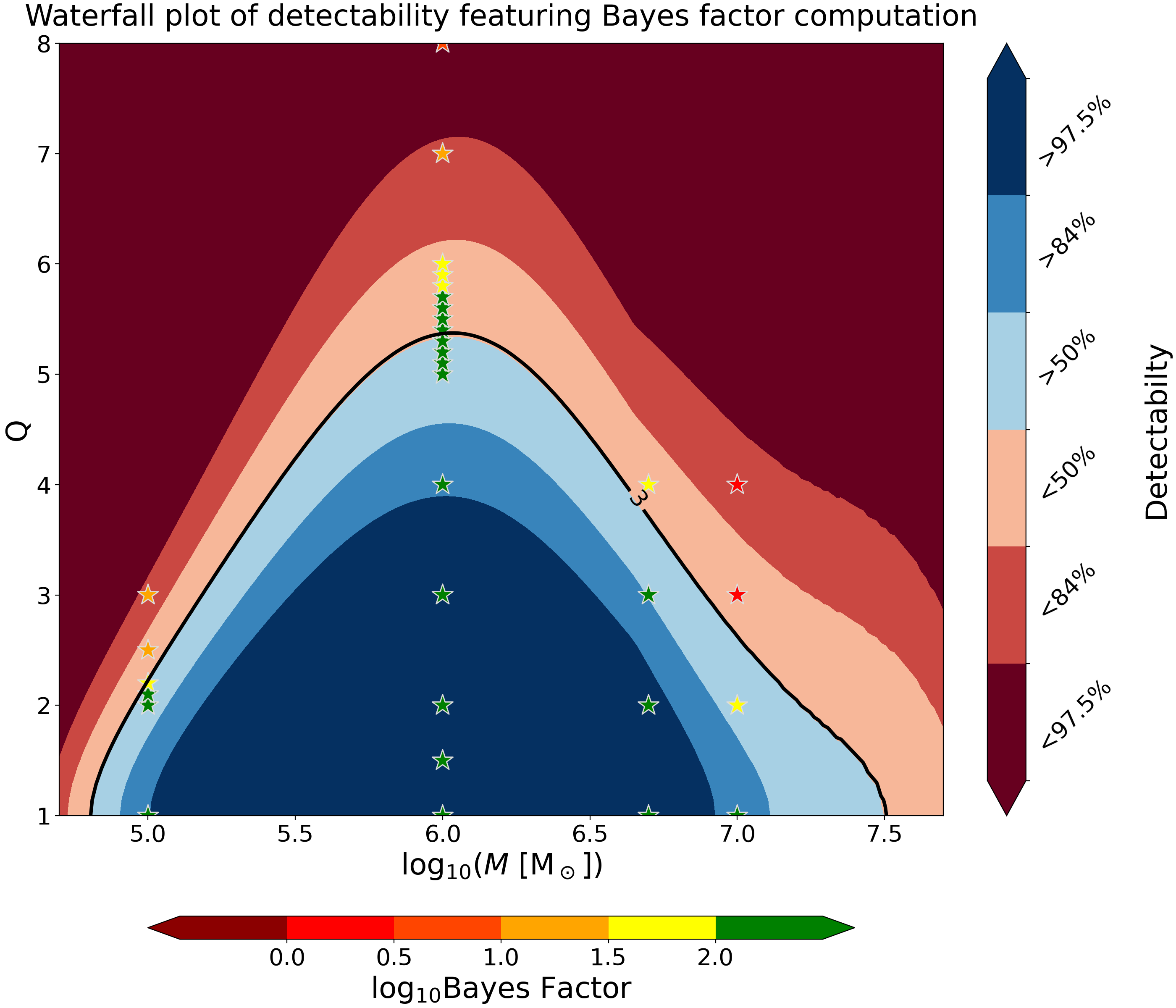}
    \caption{Conversion of the SNR waterfall in Fig.~\ref{fig:MemoryWaterfallPlotWithBF} into a waterfall detectability plot. The main colour bar (on the right) provides information on how likely it is to detect memory for a given set of parameters. The black line shows the $\SNRmem = 3$ threshold. We kept stars from the previous BF computations to compare with the prediction, using the same colour bar as in Fig.~\ref{fig:MemoryWaterfallPlotWithBF}.}
    \label{fig:DetectabilityMap}
\end{figure}

To conclude on the detectability, and make the results more visual, we use the previously described relation between the $\log_{10}\mathcal{B}$ and $\SNRmem$ to convert the waterfall plot (Fig.~\ref{fig:MemoryWaterfallPlotWithBF}) into a detectability map, where we also drew the computed $\log_{10}\mathcal{B}$ points. The result is shown in Fig.~\ref{fig:DetectabilityMap} where we explicitly show how likely it is to detect the memory effect depending on the parameters $Q$ and $M$. A similar figure, Fig.\ref{fig:WaterfallPlotRedshift}, is available in appendix to evaluate the detectability depending on redshift.

\section{Memory effect and parameter estimation}

In this section, we consider the effect of the memory effect through MBHB parameters estimation.
We use the parameter estimation that comes as a by-product of the dynamic nested sampling process we used in the previous section for estimating Bayesian posteriors and evidences~\cite{Speagle_2020}.
Two cases are considered: i) the first one is a comparison between parameter reconstruction with and without memory, to see if taking into account the memory will help to refine the results; ii) the second one is the reconstruction of the memory using ``free" geometrical parametrization via the amplitude of the memory to see how well the General Relativity prediction can be tested. 
In this section, we mostly worked using {\tt NRHybSur3dq8\_CCE} restricted to the (2,2)-mode (and memory). 
However, consistency checks have been performed and passed with {\tt SEOBNRv5HM} as well (and also a few runs of {\tt NRHybSur3dq8\_CCE} with HMs, which are much more time-consuming).

\subsection{MBHB parameter estimation with and without the memory effect}
\label{sec:ParamEstimation}
\subsubsection{Focus on the (2,2) and the main mass regime}
We first want to study if the memory effect improves the accuracy of the estimation. We performed this analysis in different scenarios and checked whether the parameter estimation is either unchanged or improved.

We need to distinguish two cases: one where the merger will be resolved by LISA, for mass $M \gtrsim 10^5 ~\mathrm{M_\odot}$, and one where the parameter estimation relies more or less on the inspiral only, $M \lesssim 10^5 ~\mathrm{M_\odot}$.
In the first case, we observe almost no difference in the parameter estimation between the two models, in particular in cases where we had high values of $\SNRtot$ and an already Gaussian, posterior distribution without memory.

More interesting effects can appear in the lower SNR cases where $\SNRtot \in [\sim100, \sim500]$. 
In this case, the posterior distributions are wider and not always Gaussian. Some parameters may be unresolved and the memory effect can then play a role, bringing additional information. In some cases for example, depending on the parameters, the noise realization can lead to degeneracies or multimode posterior distribution.

An example of such a case where we compare the results obtained with two different noise realizations is shown in Appendix, Figs.~\ref{fig:LowSNRDoubleCornerClear} and~\ref{fig:LowSNRDoubleCornerDegenerate}. Depending on the noise, we obtain either monomode posterior distributions (Fig.\ref{fig:LowSNRDoubleCornerClear}) or a multimode situation where memory allows to lift the degeneracies between parameters (Fig.~\ref{fig:LowSNRDoubleCornerDegenerate}). In the multimode case, memory offers a significant improvement in the MBHB parameter estimation, as described in previous studies~\cite{Gasparotto_2023, LVKDistanceInclDegen}.

As a first intermediate conclusion, the memory effect appears to provide useful information primarily in the low-SNR regime, as pointed out in~\cite{Gasparotto_2023}. In that work, a detailed Fisher-matrix analysis was performed, and the potential of memory to alleviate the degeneracy between inclination and luminosity distance was discussed. Building on the results of~\cite{Gasparotto_2023}, we expect the impact of memory to be more pronounced for light, equal-mass binaries, whose merger signal is largely masked by LISA noise at high frequencies and which lack a long or clearly resolved inspiral~\footnote{An interesting avenue is to consider the impact on parameter estimation for out-of-band mergers whose ringdown may fall in the sensitivity band of ground-based detectors as considered in Ref.~\cite{GasparottoGasparotto:2025adb}.}.

The impact of memory on parameter estimation discussed here is obtained mostly by including only the $(2,2)$ mode. In the more general case, the inclusion of HMs is expected to lift parameter degeneracies more efficiently than memory alone for higher masses  $M \gtrsim 10^5 ~\mathrm{M_\odot}$.

\subsubsection{HMs in the low mass regime and (2,0) oscillatory component}
\begin{figure*}
    \centering
    \includegraphics[width=0.42\linewidth]{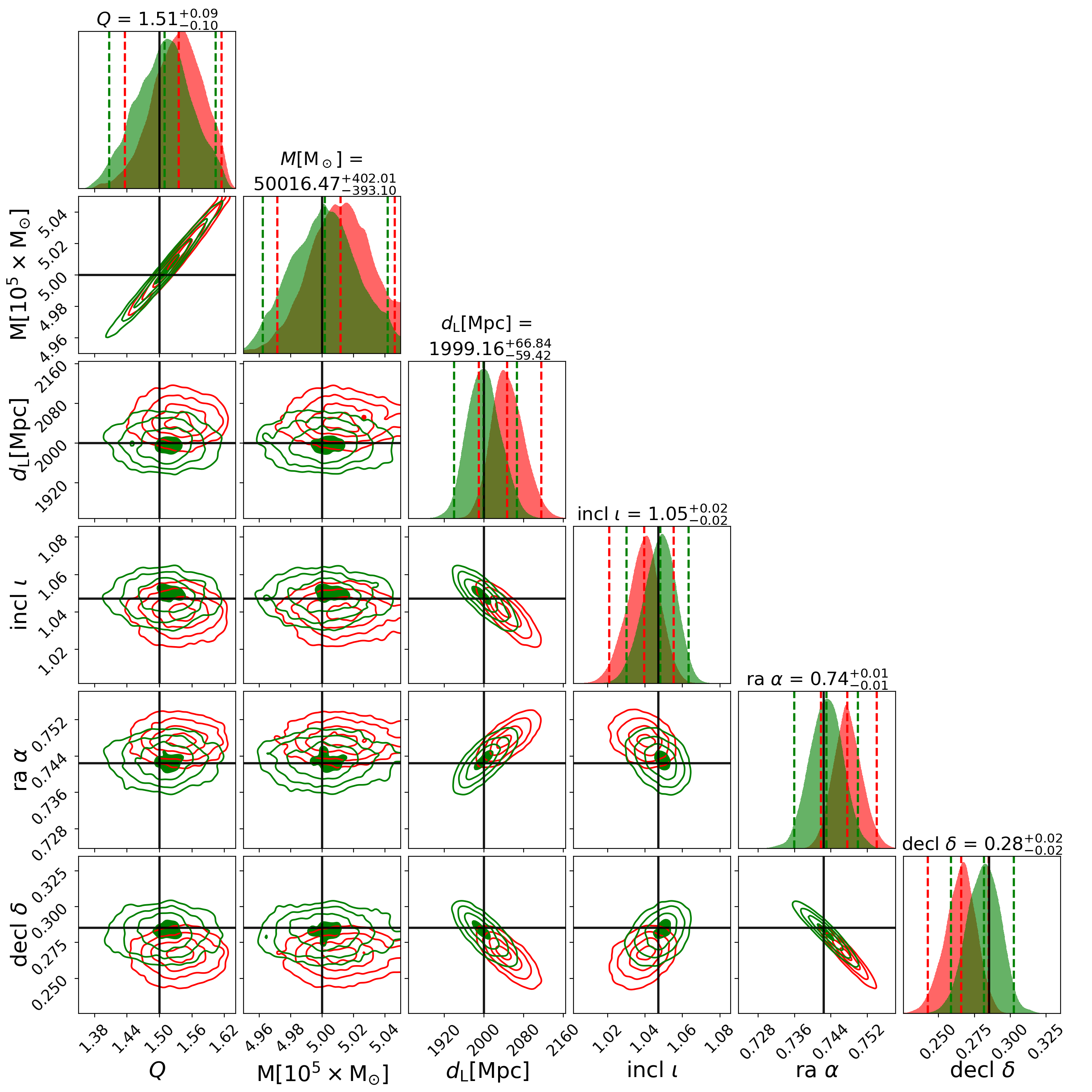}
    \includegraphics[width=0.42\linewidth]{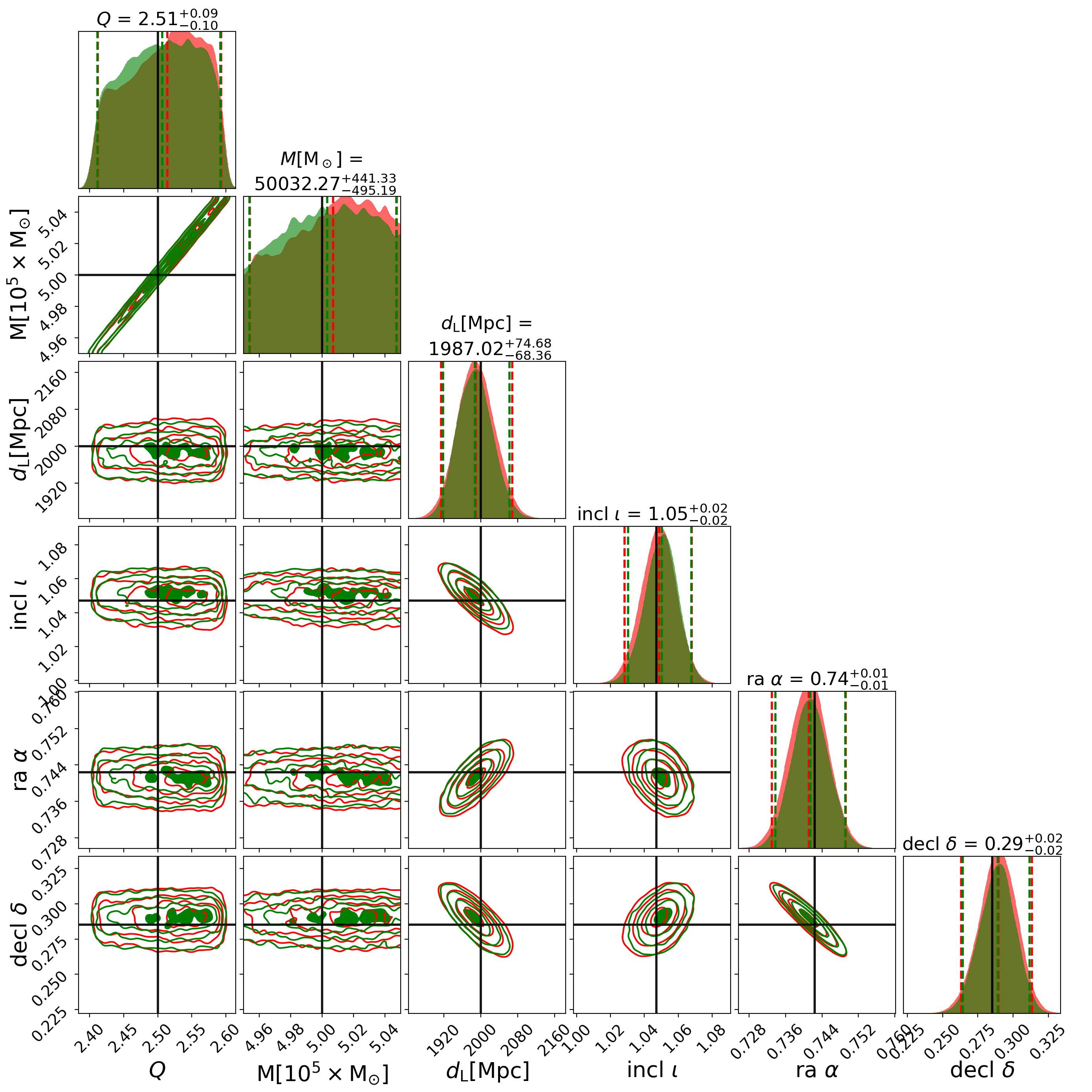}
    \caption{Cropped cornerplots showing parameters estimation of $Q$, $M$, $d_{\mathrm{L}}$, $\iota$, $\alpha$, $\delta$. We plotted the model with memory in green, and the one without in red. The values and uncertainties for parameters indicated on top of the distribution correspond to the memory model. The source parameters are $Q=1.5$ (left) / $Q=2.5$ (right) and [$\chi_{\mathrm{1z}} =\chi_{\mathrm{2z}}=0.7$, $M=5\times10^4 ~\mathrm{M_\odot}$, $d_{\mathrm{L}}=2000$ Mpc, $\iota=\pi/3$, $\varphi_{\mathrm{ref}}=1$, $\psi=0$, $\alpha=0.74$, $\delta=0.29$] Here we used the {\tt NRHybSur3dq8\_CCE} waveform with HM.}
    \label{fig:DoubleDoublecornerplots}
\end{figure*}

However, in the low-mass regime, $M \lesssim 10^5 ~\mathrm{M_\odot}$, the information from the merger -- and in particular from HMs with $\ell \geqslant 3$ -- lies predominantly in a frequency range where LISA is less sensitive, thereby reducing the quality of the reconstructed information. In this case, the low-frequency nature of the memory signal may help to improve parameter constraints.
We performed additional runs with $M = 5 \times 10^4 ~\mathrm{M_\odot}$ and HMs. Among them we obtained a variety of outcomes. In some cases, both models -- with and without memory -- accurately reconstruct the injected parameters; in others, the inclusion of memory helps to mitigate a bias present in the model without memory; finally, there are cases in which both models exhibit similar biases. Fig.~\ref{fig:DoubleDoublecornerplots} shows an example of the two first cases. Overall, these results suggest that memory is not decisive for achieving clean parameter estimation in the general case, although it can be beneficial in specific situations. Further work is required to clearly identify the conditions under which memory becomes a key ingredient for parameter estimation.

At this stage, it is also worth mentioning the effect of including the full $(2,0)$ mode, which offers interesting insights into the relative role of the memory effect compared to higher-order modes, as discussed in Rossello et al.~\cite{Rossello_2025}. In the time domain, the amplitude of the oscillatory component of the $(2,0)$ mode is typically small compared to that of the memory signal. However, owing to LISA’s response, the oscillatory component can still dominate over the more step-like memory signal, potentially leading to significant biases in parameter estimation (see~\ref{app:20oscill}). This highlights the importance of incorporating HMs in parameter estimation, which may be more impactful than including the memory effect alone, although the latter can still provide a useful additional handle.

\subsection{Estimating the memory's amplitude}

\begin{figure}[h!]
    \centering
    \includegraphics[width=0.9\linewidth]{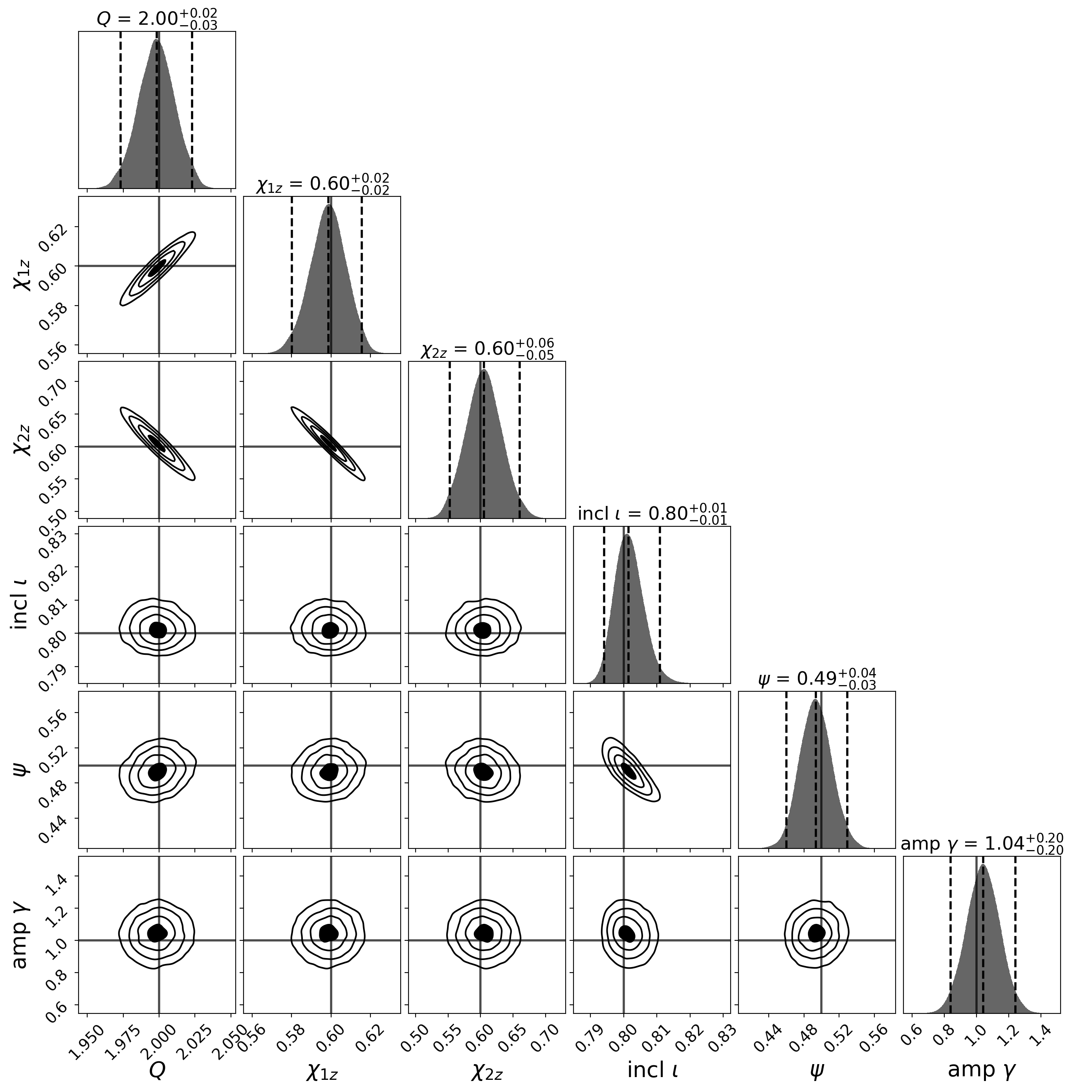}
    \caption{Cropped corner plot showing parameters estimation from mock data with GR memory --i.e. $\gamma = 1$--, using a model with memory where the amplitude of the latter is parametrized. {\tt NRHybSur3dq8\_CCE} is used. Injection parameters are located with the black lines.}
    \label{fig:CroppedCornerWithAmp}
\end{figure}

We now aim to measure the shape of the memory burst.
We start limiting ourselves to evaluate the amplitude, which is the primary characteristic of the memory burst.
One can indeed expect that the amplitude of a memory burst will likely be the first parameter to be affected in the case of a deviation from GR.

To control the amplitude, we add a parameter $\gamma$ defined as $h_{\mathrm{mem}}(t) = \gamma \times h_{\mathrm{mem}}^{\mathrm{GR}}(t)$ where $h_{\mathrm{mem}}$ is the modified memory waveform and $h_{\mathrm{mem}}^{\mathrm{GR}}$ is the original GR memory computed using Eq.~\eqref{eq:BMS_BalanceLaw}. 

Fig.~\ref{fig:CroppedCornerWithAmp} shows an example of the posteriors when this parametrization of the memory effect is included. 
One can see that the amplitude of the memory effect can be quite well reconstructed albeit with significant uncertainty.
This analysis on the shape of memory is consecutive to the analysis previously detailed in Section~\ref{sec:ParamEstimation}. We apply this to mock datasets in which we have a clear detection of the memory effect, i.e. $\log_{10}\mathcal{B} \gtrsim 3$. We therefore suppose  $\SNRmem \gtrsim 5$ and a narrower prior distribution, as we can use the information gathered in the previous parameter estimation.

\begin{figure}[H]
    \centering
    \includegraphics[width=1\linewidth]{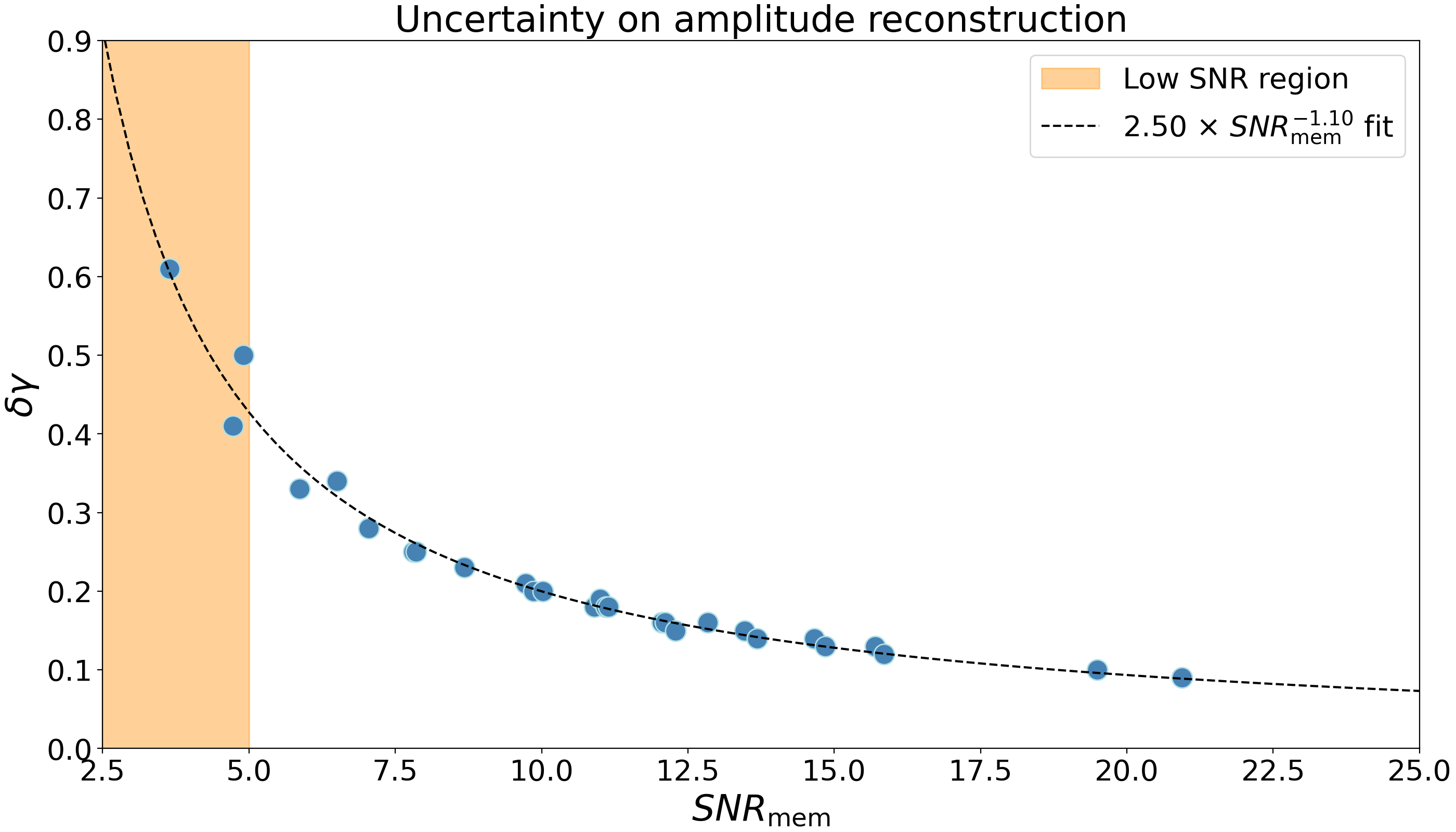}
    \caption{Measured uncertainties on the amplitude $\delta \gamma$ depending on the $\SNRmem$ of the input GW. 
    Every point is a measurement using a different set of parameters and noise realization. 
    The yellow area indicates the points where the detectability of the memory is likely, yet uncertain due to noise.
    The dashed line is the fit by a power law.}
    \label{fig:DeltaAmplitude}
\end{figure}

The next step is to monitor the uncertainty on the reconstructed amplitude $\delta \gamma$. 
To do this, we run the dynamic nested sampling package on various sets of parameters and then check for correlation between the uncertainty $\delta \gamma$ and binary parameters (namely all the intrinsic binary parameters, the inclination and both $\SNRtot$ and $\SNRmem$). 
We found a direct link between $\delta \gamma$ and $\SNRmem$, which is shown in Fig.~\ref{fig:DeltaAmplitude}.
The values of $\delta \gamma$ follow a power-law in $\SNRmem$, namely 
\[\delta \gamma = 2.5 \times \SNRmem^{-1.1},\]
which allows us to predict the $\SNRmem$ required to achieve a given uncertainty and, consequently, to identify the class of sources needed to reach a desired precision in the amplitude reconstruction. Considering the uncertainty on the power, we have $1.10 \pm 0.03$, highlighting a slight deviation from a simple linear behaviour.

\begin{table}[H]
 \begin{ruledtabular}
    \begin{tabular}{cc}
        Targeted uncertainty & $\SNRmem$ required \\
        \hline
         25\% & 8.15 \\
         20\% & 9.99 \\
         15\% & 13.0 \\
         10\% & 18.8 \\
         5\% & 35.4 \\
    \end{tabular}
    \caption{Using the power law linking $\delta \gamma$ and $\SNRmem$, namely $\delta \gamma = 2.5 \times \SNRmem^{-1.1}$,  one can estimate the minimal $\SNRmem$ value needed to achieve a certain precision on the amplitude measurement.}
    \label{tab:UncertaintyOnAmplitude}
 \end{ruledtabular}
\end{table}
Table~\ref{tab:UncertaintyOnAmplitude} shows numerical values of $\SNRmem$ needed to reach a given precision on the amplitude's measurement.
We use this criterion with a population model, as we will see in Section~\ref{sec:Catalogs}. In that case, we can estimate how precisely we can test GR and constrain alternative models predicting a change of amplitude for the memory effect.

Pushing this analysis further using a geometrical, model-agnostic template for the memory effect will enable capturing beyond-GR features in the signal. Hence, one could allow for modifications of amplitude and the time scale of memory growth parameters around the merger, for example. Indeed, these two parameters further define the shape of the memory strain and can serve both as a test and as a trigger for possible beyond GR theories. This is an agnostic approach to test GR in the framework of LISA. It can be completed considering other detectors such as Einstein Telescope \cite{Goncharov_Donnay_Harms_2024}. More information on the effect of beyond-GR theories on the memory effect can be found, for example, in~\cite{Garfinkle_Hollands_2017, Hollands_2017, Heisenberg_2023, Heisenberg_2025}. We also aims to use a more complete model for memory, not reduced to its component in the $(2,0)$ mode. This is left aside for future work.

\section{Comparison with catalogues}
\label{sec:Catalogs}
\subsection{General method}

\begin{table*}
 \renewcommand{\arraystretch}{1.3}
 \begin{ruledtabular}
    \begin{tabular}{ccccccc}
        \multirow{3}{*}{SN} & \multirow{3}{*}{Delays} & \multirow{3}{*}{Seeds} & Nb of events & Nb of events & Nb of events & Nb of events \\[-1ex]
        & & & detected & with & with & to reach \\[-1ex]
        & & & ($\SNRtot > 8$) & $\SNRmem>3$ &  $\SNRmem>5$ & $\log_{10}B^{cumul} > 2$\\
        \hline
         \multirow{4}{*}{Yes} & Short & Light & 0.0$^{+0.0}_{-0.0}$ & 0.0$^{+0.0}_{-0.0}$ & 0.0$^{+0.0}_{-0.0}$ & N\textbackslash A \\ 
                            \cline{3-7}
                            & delays & Heavy & 1024$^{+46}_{-47}$ & 87$^{+16}_{-15}$ & 37$^{+10}_{-10}$ & 7.0$^{+0.7}_{-0.6}$\\
                            \cline{2-7}
                            & \multirow{2}{*}{Delays} & Light & 0.0$^{+0.0}_{-0.0}$ & 0.0$^{+0.0}_{-0.0}$ & 0.0$^{+0.0}_{-0.0}$ & N\textbackslash A\\
                            \cline{3-7}
                            &           & Heavy & 21$^{+8.0}_{-6.0}$ & 11$^{+6.0}_{-5.0}$ & 8.0$^{+5.0}_{-4.0}$ & 1.7$^{+0.7}_{-0.4}$ \\
        \hline
         \multirow{4}{*}{No} & Short & Light & 38.0$^{+10}_{-10}$ & 2.0$^{+3.0}_{-2.0}$ & 1.0$^{+2.0}_{-1.0}$ & 10.8$^{+8.7}_{-4.5}$ \\
                            \cline{3-7}
                            & delays & Heavy & 1033$^{+48}_{-52}$ & 81$^{+16}_{-15}$ & 32$^{+10}_{-9.0}$ & 7.3$^{+0.8}_{-0.7}$ \\
                            \cline{2-7}
                            & \multirow{2}{*}{Delays} & Light & 13.0$^{+6.0}_{-6.0}$ & 1.0$^{+3.0}_{-1.0}$ & 1.0$^{+2.0}_{-1.0}$ & 5.2$^{+4.6}_{-2.2}$\\
                            \cline{3-7}
                            &           & Heavy & 8.0$^{+5.0}_{-4.0}$ & 4.0$^{+3.0}_{-3.0}$ & 3.0$^{+3.0}_{-3.0}$ & 1.8$^{+1.4}_{-0.6}$
    \end{tabular}
    \caption{Table with the expectation of detection for different models, using Barausse's populations. The values are provided as $\textrm{median}^{+2\sigma}_{-2\sigma}$ to reflect the asymmetries of the distributions.}
    \label{tab:BarausseCatalogsExpectation}
 \end{ruledtabular}
\end{table*}

To put our results in perspective, we use the recent MBHB catalogues from Barausse et al.~\cite{Barausse_2020, Barausse_Lapi_2021}, also used in~\cite{Gasparotto_2023,Henri_Memory_Paper}, to estimate the number of MBHB we expect under different scenarios and if their associated memory can reach the $\SNRmem$ threshold.

This catalogues provide MBHB parameters such as masses, spins, and redshifts, grouped into 1-year data files.
Sky position, inclination, phase, and polarization are randomly drawn, assuming uniform distributions\footnote{Except for the inclination $\iota$, where $\cos(\iota)$ is drawn uniformly, representing isotropically inclined MBHs.}.
For each model, we simulate 1000 realizations of 4-year observations—the nominal LISA mission duration~\cite{LISA_Redbook}—by randomly combining 1-year files. For every binary, we compute the GW emission and the total and memory signal-to-noise ratios, $\SNRtot$ and $\SNRmem$. We also generate 1000 realizations for 10-year observations, corresponding to LISA’s maximum mission extension.

The SNR computation is done using the {\tt NRHybSur3dq8\_CCE} waveform and taking its higher modes into account, as computing the SNR is fast enough. However, the {\tt NRHybSur3dq8\_CCE} waveform does not include the BH precession, whose absence, we remind the reader, is a simplifying assumption throughout this work. 
In consequence we flatten the spin components by keeping only the $z$-direction that is along the orbital momentum (e.g. if a catalogue entry indicates a binary with $(\vec{\chi_1},\vec{\chi_2}) = ([0.1, 0.2, 0.8], [0.4, -0.2, 0.6])$, we keep it as $(\vec{\chi_1},\vec{\chi_2}) = ([0, 0, 0.8], [0, 0, 0.6])$).
In addition, the {\tt NRHybSur3dq8\_CCE} waveform does not allow the computation of MBHB waveforms with $Q > 10$ and therefore we discard them from the analysis. This should have a limited impact on the results as there are few events with $Q>10$ in the catalogue, and their memory is highly suppressed as suggested from Fig.~\ref{fig:DetectabilityMap}.
Given these assumptions, the reliability of these predictions depends on how realistic the catalogues are and, in particular, on the impact of neglecting precession. If all memory modes could be fully separated, precession might actually enhance detection and characterization, as it introduces additional memory components in the $m=1$ modes~\cite{GWmemory}.

From this SNR computation, we can use our previous results to estimate how many binaries will have observable memory effect, i.e. $\SNRmem > \SNRmem^{\textrm{thresh}} = 3$, or $\SNRmem > 5$ which ensures detectability most of the time, regardless of any noise realization.

Finally, we confirm that the memory effect is likely observable for some high–SNR sources, although its detection may require accumulating data over time. To assess its detectability, we perform a cumulative Bayesian analysis following~\cite{Cheung_2024, ThanksForTheMemory, MemoryRemains}. For each model, we randomly select events in every universe realization with $\SNRtot > 8$, as in~\cite{Barausse_2020, Barausse_Lapi_2021}, and determine the number of such events required to exceed the detection threshold $\log_{10}\mathcal{B}^{\mathrm{cumul}} > 2$, where $\mathcal{B}^{\mathrm{cumul}}$ is the cumulative Bayes factor (Eq.~\eqref{eq:CumulativeBayes}).

\begin{equation}
    \mathcal{B}^{\mathrm{cumul}} = \prod_i \mathcal{B}_i \ \Rightarrow \ \log_{10}\mathcal{B}^{\mathrm{cumul}} = \sum_i \log_{10}\mathcal{B}_i
    \label{eq:CumulativeBayes}
\end{equation}
This is done 500 times to obtain a mean value for each model.

\subsection{MBHB population catalogues}
\label{sec:BarausseCat}

Population catalogues introduced by Barausse et al.~\cite{Barausse_2020, Barausse_Lapi_2021} present 8 different models. 
These models have three main characteristics, namely seeding, delays, and supernovae feedback. The seeding corresponds to the underlying hypothesis on the formation of massive black holes (MBH).
The ``light-seed" scenario assumes the formation of intermediate-mass black holes (IMBH) from remnants of Population III stars in the early universe and mergers between these remnants to form IMBHs and MBHs. The ``heavy-seed" scenario assumes the formation of IMBHs and MBHs up to $10^6 ~\mathrm{M_\odot}$ in the early universe through mechanisms like direct collapse of large gas clouds, before additional mergers as in the light-seed scenario.
The second characteristic is the delay. MBHs are usually hosted at galaxy centers. During the merging process, MBHs interact with the background environment before merging, which can imply a delay between the merger of the host galaxies and the merger of the MBH. 
Lastly, these models can take the effect of supernovae (SN) feedback into account. Indeed, SN affect the ability for MBHs to grow, in particular for MBHs hosted in low-mass galaxies. 
The 8 models represent all the possibilities of mixing between the type of seed -- i.e., light or heavy --, the delays -- i.e., short delays or standard delays -- and the effect of SN -- i.e, taken into account or not --.

Table~\ref{tab:BarausseCatalogsExpectation} shows a summary of the results obtained with these models. One can notice that two of the eight models -- i.e. the ones with light seed and supernovae feedback, independently of the delay -- predict no visible event (considering the threshold at $\SNRtot >8$ used in~\cite{Barausse_2020}) as their predicted MBHBs have mostly low total mass $M \lesssim 10^4 ~\mathrm{M_\odot}$. In~\cite{Barausse_2020}, Barausse et al. predict more visible sources than in our work. This can be explained by the fact that we only consider the late inspiral, the merger, and the ringdown, as we focus on the memory effect. Additional sources may be observed during their inspiral phase, but they will not aid in memory detection. These two models will not appear in the following figures for clarity.

\begin{figure}
    \centering
    \includegraphics[width=0.9\linewidth]{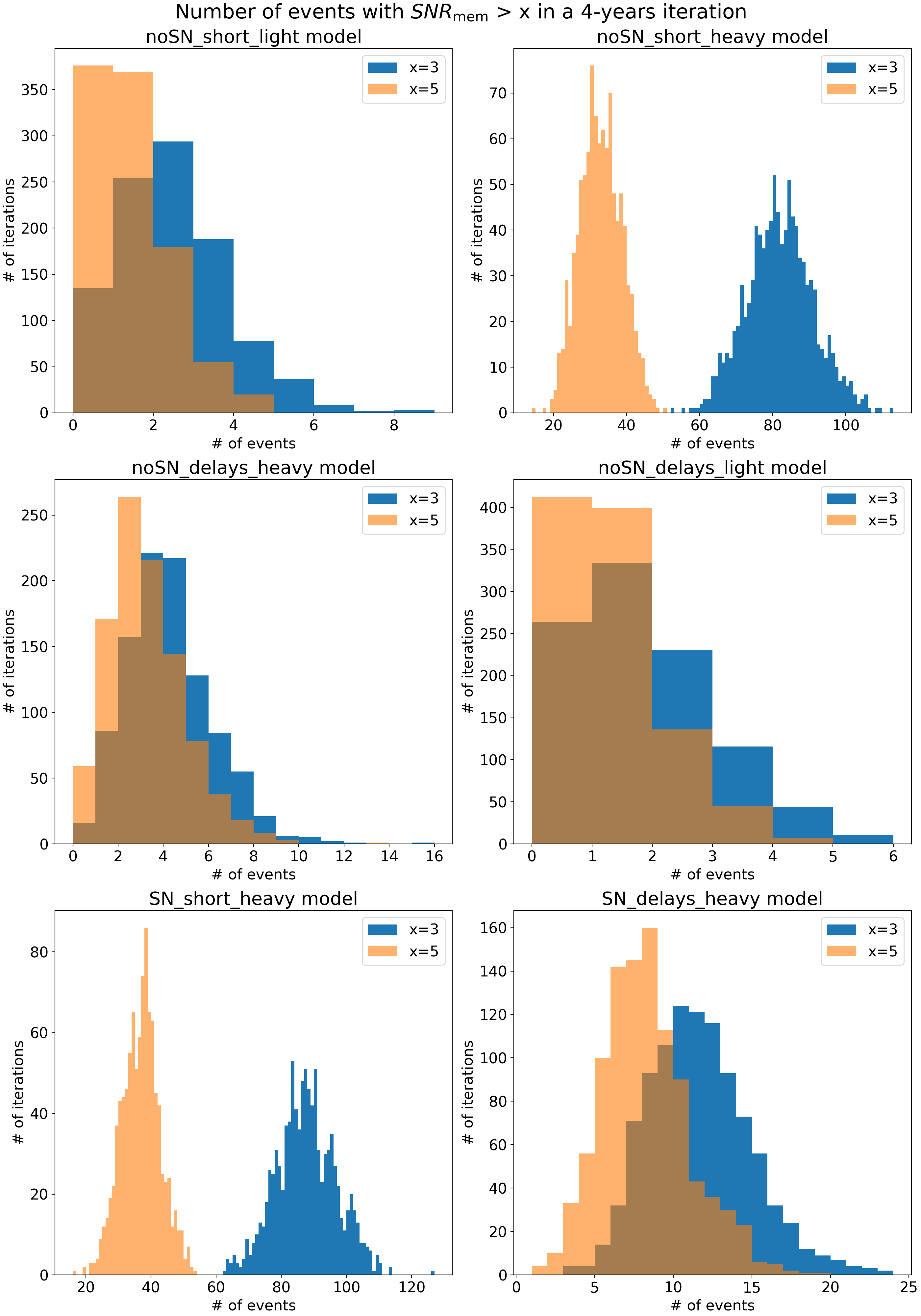}
    \caption{Histograms of the number of events such that $\SNRmem > 3$ (blue) and $\SNRmem > 5$ (orange) for the six remaining Barausse's catalogues. The bins are unitary and each model presents realizations of 4-year data. A 10-year version can be found in the appendix, Fig.~\ref{fig:Barausse_10yrs_MemorySeenSources}.}
    \label{fig:Barausse_4yrs_MemorySeenSources}
\end{figure}

Fig.~\ref{fig:Barausse_4yrs_MemorySeenSources} shows the probability distribution of the number of memory events seen in four years of observation. 
The relative plot for 10 years of observation is provided in the appendix, Fig.~\ref{fig:Barausse_10yrs_MemorySeenSources}.

To estimate our ability to test GR, it is interesting to look at the highest $\SNRmem$ value reached in a universe realization. 
For each population model, we can consider a realization $i$, and define its maximum value as $\SNRmem^{\mathrm{max}}(i)$. Considering $N = 1000$ the number of all realizations per population model, and $N_x$ the number of realizations such that $\SNRmem^{\mathrm{max}}(i) > x$, we can express the probability of having at least one event with $\SNRmem > x$ in a realization as $P(x) = \frac{N_x}{N}$.

\begin{figure}[h!]
    \centering
    \includegraphics[width=0.9\linewidth]{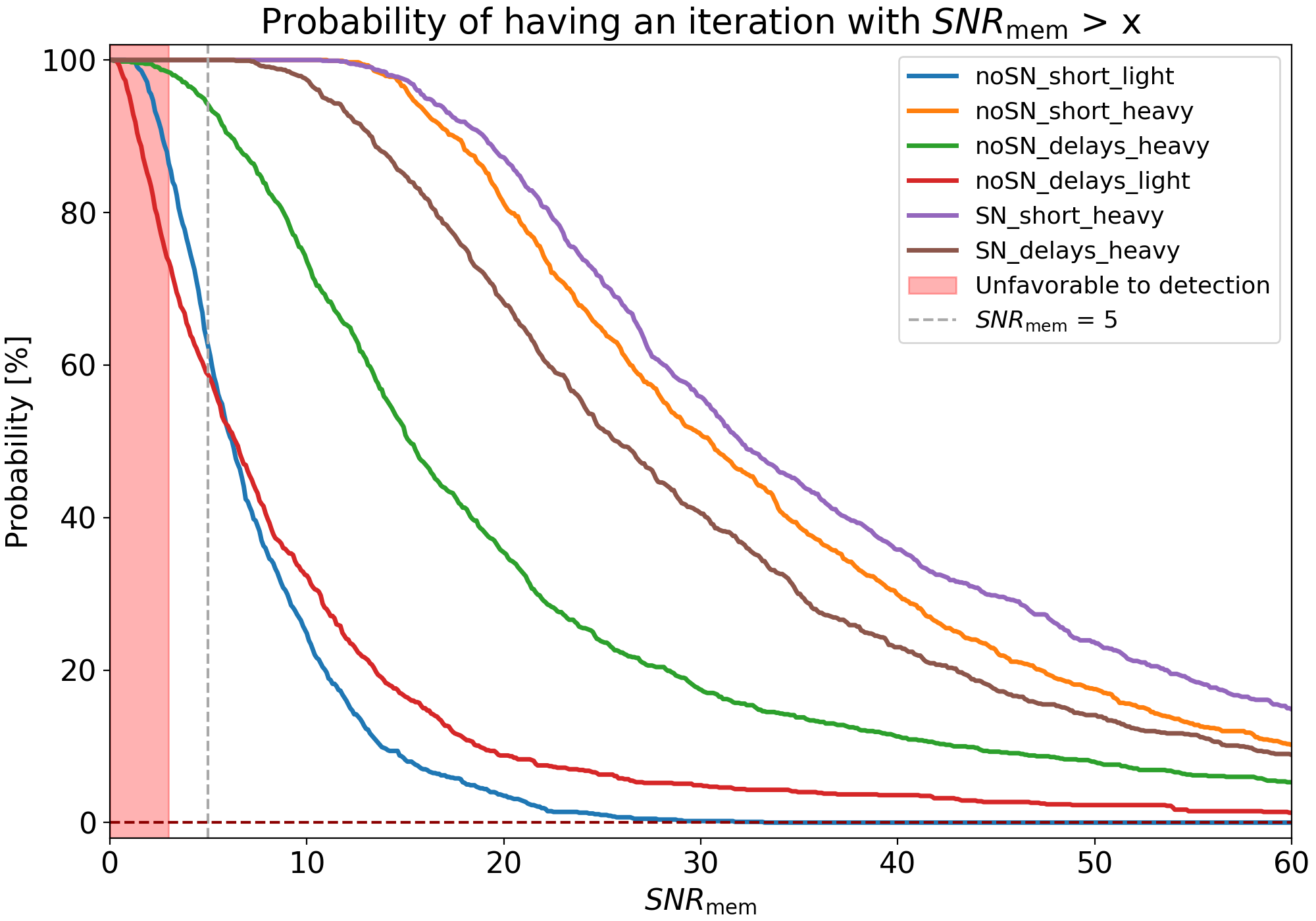}
    \caption{Probability of having a 4-years iteration with $\SNRmem$ greater than a given value (x-axis). Each solid line corresponds to a population model from Barausse et al.~\cite{Barausse_2020, Barausse_Lapi_2021}. The red area cover the region where we are under the threshold $\SNRmem^{\textrm{thresh}} = 3$. The gray dashed line shows the value $\SNRmem = 5$ over which memory should be always detected. A 10-year version can be found in the appendix, Fig.~\ref{fig:Barausse_10yrs_ProbaOfMax}.}
    \label{fig:Barausse_4yrs_ProbaOfMax}
\end{figure}

Table~\ref{tab:BarausseMaxSNRProba}, in appendix, summarizes the probability of having an realization with at least one event with an $\SNRmem$ bigger than a given reference $\SNRmem$. Fig.~\ref{fig:Barausse_4yrs_ProbaOfMax} offers a visual approach to these results. Fig.~\ref{fig:Barausse_10yrs_ProbaOfMax}, in the appendix, provides a complementary view with the 10-year observation results.

In conclusion, heavy-seed scenarios provide promising prospects for detecting memory bursts, whereas light-seed scenarios show greater variability but may still yield favourable events. Moreover, recent studies such as~\cite{Barausse_2023}, which incorporate PTA evidence for the stochastic GW background, suggest shorter delays and favour models predicting a higher number of detectable events.

\section{Conclusion \& discussion}
\label{sec:Conclusion}

In this paper, we pushed a step further the work from Inchauspé et al.~\cite{Henri_Memory_Paper} on the study of the displacement memory effect, and in particular its main component in the $(2,0)$ mode. Our first objective was to set up a Bayesian analysis to confirm the detectability of the memory effect for different sets of MBHB parameters. Our results confirmed the expectation from~\cite{Henri_Memory_Paper} that, in some cases, the displacement memory effect is sufficiently loud to be detected. From this, we were able to establish a link between the Bayes factor $\mathcal{B}$, comparing the models with and without memory, and the SNR of the memory effect, $\SNRmem$, leading to the definition of a threshold in $\SNRmem$ where detection starts to be possible, namely $\SNRmem^{\textrm{thresh}} = 3$. We also found that, as expected, the Bayes factor will be affected by the LISA instrumental noise, and we characterized the induced dispersion due to this noise to take it into account in our detectability study. If this dispersion depends on the total mass of the binary, in the detector frame, we can still establish that, in most cases, GWs with an $\SNRmem > 5$ should allow confident detection of the memory effect. These results enable us to establish a map of the parameter space, indicating where, and with what confidence, one can detect the memory effect.\\
Next, we investigate the impact of the memory effect on the MBHB parameter estimation. One often considers high $\SNRtot$, where we showed that the memory effect mostly does not affect the MBHB parameter estimation, as it is already highly constrained by the oscillatory component of the GW. However, for quieter signals, $\SNRtot \sim 100$, the memory effect can help lift some parameter degeneracies, such as the luminosity distance $d_{\mathrm{L}}$ and the inclination $\iota$, or between sky coordinates. In these cases, our results are in agreement with previous work such as~\cite{Gasparotto_2023}. We leave for future work a systematic analysis of the parameter-estimation biases that may arise from neglecting the memory contribution, similarly to~\cite{Pitte:2023ltw} for the other higher harmonics. 
\\
Having shown that we can characterize the MBHB parameters, we investigated whether we could reconstruct the geometrical parameters of the memory effect, such as the amplitude during the merger. 
We found that the amplitude can be measured with reasonable precision without affecting much the quality of the parameter estimation. In particular, we have shown that the precision of the measurement of the amplitude of the memory effect at the merger can be predicted as a function of the $\SNRmem$ of the signal. 
This enables an assessment of LISA’s sensitivity to the amplitude of the memory effect, which can then be compared with the predictions of GR.
A more comprehensive study, including more parameters, such as the time scale of the memory effect rise, would be interesting to perform to further test GR. 
This study is left for future work.\\
Last but not least, we used a catalogue of simulated MBHB populations to place our results in an astrophysical context. Despite the uncertainties on the catalogue itself and despite some of our work assumptions, hypotheses, and intrinsic limitations of the waveform we used, this study improves previous estimates on the likelihood of observing the gravitational-wave memory effect with LISA.
By calculating the $\SNRmem$ of the sources in the catalogue, we find that LISA is expected to detect the memory effect in individual events in heavy-seed scenarios, while there is a good chance of detection also in light-seed scenarios. Within this astrophysical framework, our results therefore provide initial insights into LISA’s ability to test General Relativity through measurements of the memory effect. 
Possible future directions of this work include a more refined identification of potential ``orphan'' memory cases~\cite{Gasparotto:2025wok, McNeill:2017uvq}, as well as the inclusion of precessing binary systems. In the latter case, it will be necessary to consistently account all the different modes that includes the memory component. The analysis could also be extended to other astrophysical populations, as well as to the linear memory generated by hyperbolic encounters. Ultimately, the memory effect should be incorporated into a full global fit analysis.

\begin{acknowledgments}
    We would like to thank all the members of the LISA memory working group for the useful discussion and feedback that has helped the project to progress. We would also like to thank E. Barausse for granting us access to the MBHB population catalogues on which he worked.
    This work was supported by CNES, focused on LISA Mission. AC acknowledges financial support from the Commissariat à l'Énergie Atomique (CEA) and from the Centre national d’études spatiales (CNES), within the framework of the LISA mission. JZ is supported by funding from the Swiss National Science Foundation
    (Grant No. 222346) and the Janggen-Pöhn-Foundation. The Center of Gravity is a Center of Excellence funded by the Danish National Research Foundation under Grant No. 184. HI thanks the Belgian Federal Science Policy Office (BELSPO) for the provision of financial support in the framework of the PRODEX Programme of the European Space Agency (ESA) under contract number PEA4000144253. CP acknowledges the Agenzia Spaziale Italiana (ASI), Project n. 2024-36-HH.0, "Attività per la fase B2/C della missione LISA”. LMZ is supported through Research Grants No. VIL37766 and No. VIL53101 from Villum Fonden and the DNRF Chair Program Grant No. DNRF162 by the Danish National Research Foundation.

    Computations were performed on the IRFU's cluster Feynman, CEA, France.
\end{acknowledgments}

\newpage
$~$
\newpage

\appendix
\begin{widetext}
\section{LISA Response to memory}\label{App:LisaResponse}

As mentioned in Sec.\ref{sec:MemLISA_waveformResponse}, we show an example of waveform with memory within the LISA TDI response. Fig.\ref{fig:TimeTDIA} shows how the TDI-A channel looks like for this input in both time and frequency domain. We distinguish the full signal (oscillatory+memory) and the memory signal alone to identify the memory imprint. A more detailed discussion on the LISA response to memory can be found in the companion paper~\cite{Zosso:2025memory}.

\begin{figure}[H]
    \centering
    \includegraphics[width=0.46\linewidth]{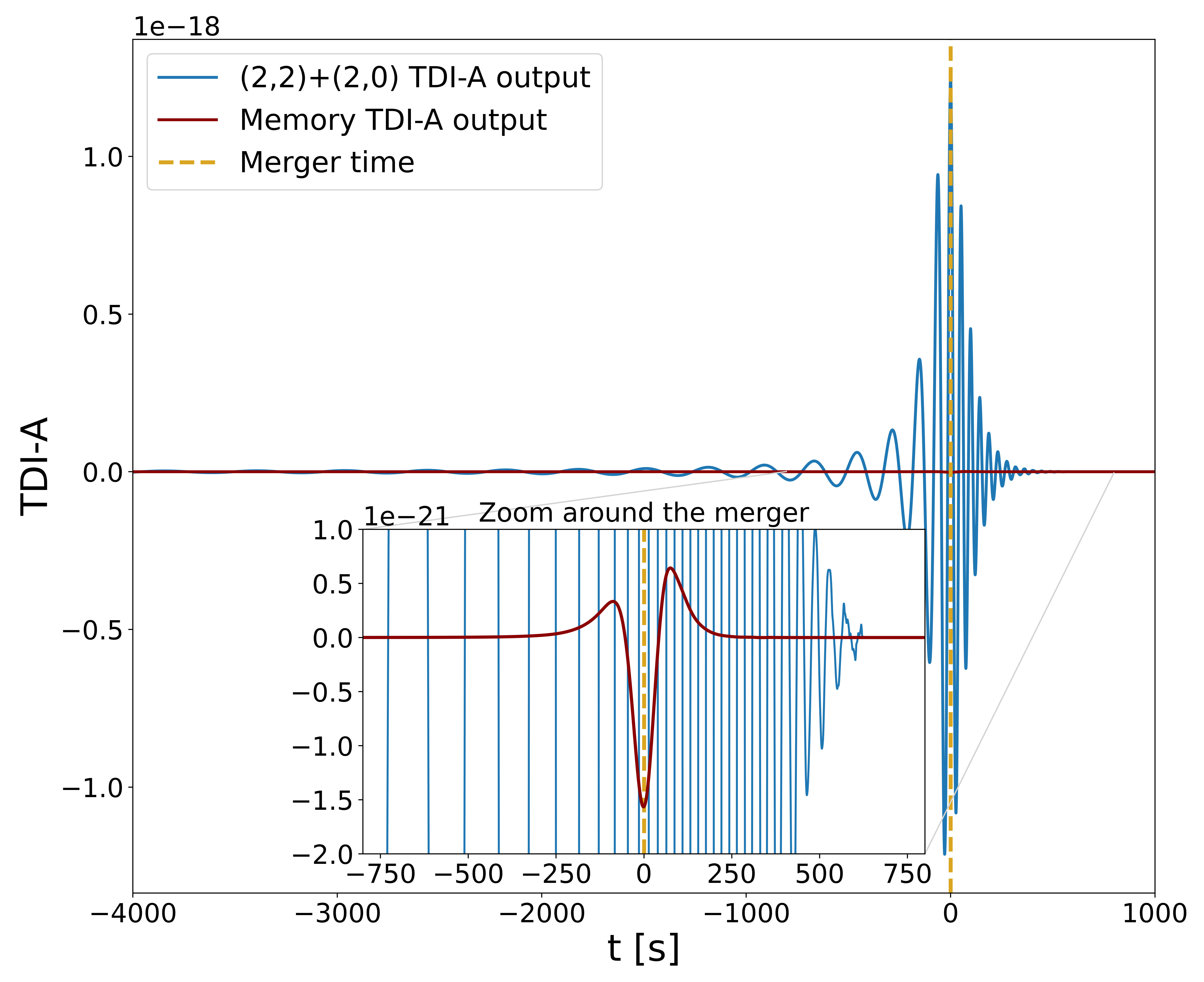}
    \includegraphics[width=0.48\linewidth]{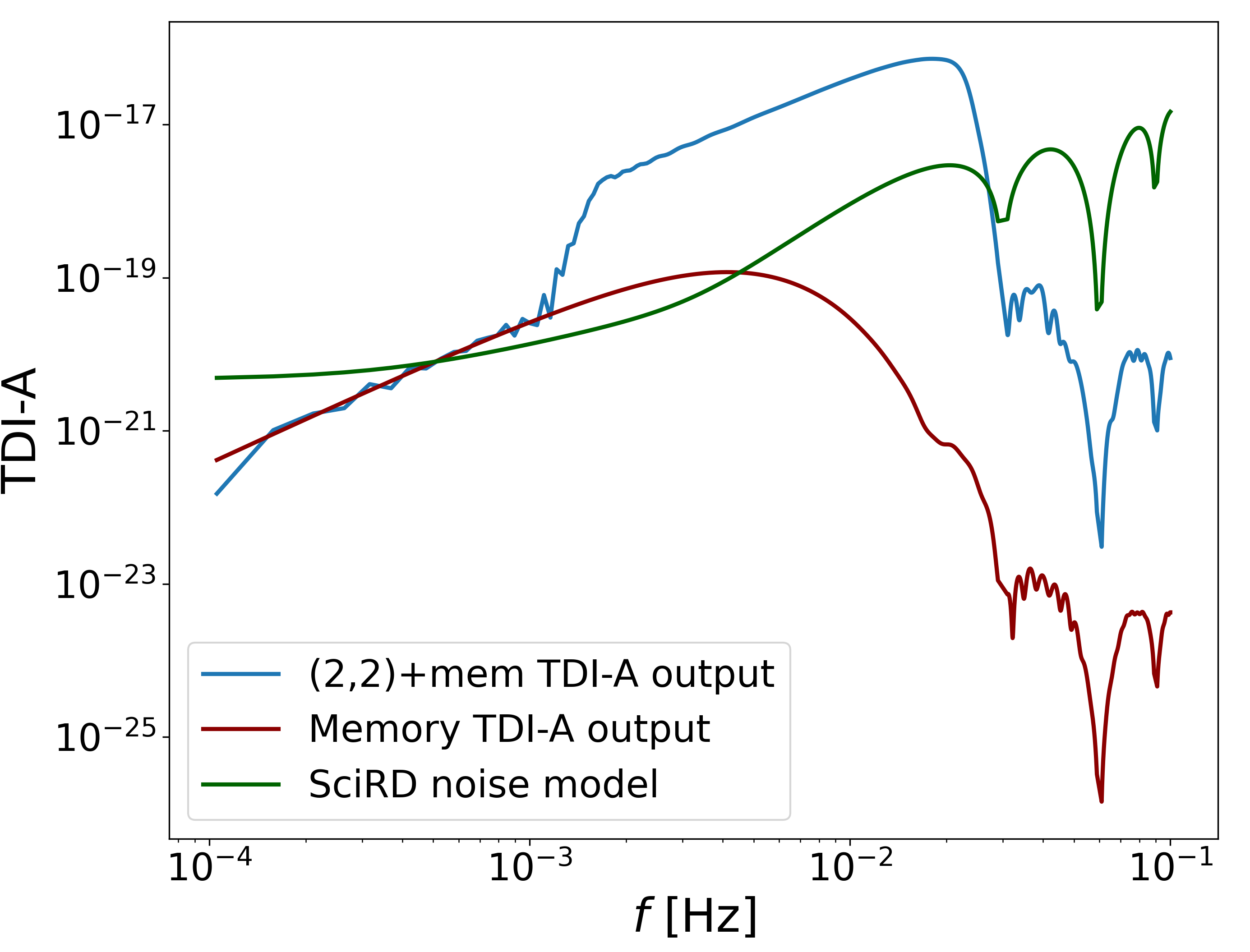}
    \caption{Time-domain (left) and frequency-domain (right) TDI-A channel obtained, after the response of the links, from the waveform illustrated in Fig.~\ref{fig:WaveformWithMem}. The total waveform is in blue and the memory component alone in red. On the left figure, the inset plot provides a clearer view of the resulting memory component. On the right figure, we added the analytical PSD of the SciRD~\cite{LISA_SciRD} noise model, in green, as a reference. This highlights the possible visibility of the memory component in the mHz region. The parameters are the same as in Fig.~\ref{fig:WaveformWithMem}, with additional sky coordinates $\alpha = 0.74$, $\delta = 0.29$.
    }
    \label{fig:TimeTDIA}
\end{figure}

In a second time, we'd like to mention some technical details on processing the signal through the response of LISA and TDI processing to ensure results reproducibility. When using a time-domain waveform with memory, one can notice that a glitch arise in the links response (not shown here) and affecting the results. This is due to the non-zero ending time-series. As the position of the glitch is located close to the end of the signal, we pad the waveform with its final value so the glitch is far away from the signal and can be cut without loss of information. We can also quickly remind that the process to correctly build TDI channels at a given instant $t$ requires information from earlier time (see~\cite{PyTDI}), therefore we cut the early response where {\tt pyTDI} don't have enough information to correctly build the TDI channels.
In the frequency-domain plot of Fig.~\ref{fig:TimeTDIA}, one can notice that both signals present unphysical features above $f \sim 2.5\times10^{-2}$ due to artefacts. This should not impact the analysis, but we kept it as it's technically taken into account in the likelihood.

\newpage
\section{Complementary waterfall plots}
\makeatletter\def\@currentlabel{Appendix B}\makeatother
\label{app:Waterfalls}

We construct waterfall plots using the {\tt SEOBNRv5HM} waveform where the SNR is computed with the same parameter values as for the {\tt NRHybSur3dq8\_CCE} case in Fig.~\ref{fig:WaterfallPlotsSurrogate} for the sake of consistency and comparison and for the same range in mass ratio $Q$ and total mass $M$.
The results are shown in Fig.~\ref{fig:WaterfallPlotsSEOBNR} where we have included HMs as described above in section~\ref{sec:MemLISA}. We also computed a waterfall plot where we have restricted the waveform to the (2,2)-mode to check the consistency between the two models of waveform, i.e., {\tt NRHybSur3dq8\_CCE} and {\tt SEOBNRv5HM}, when put in the same condition. We obtain close results for both waveforms, showing no systematic difference due to the used waveform.
The impact of subdominant modes can be seen in Fig.~\ref{fig:WaterfallPlotsSEOBNR} where, for the total SNR plot, a slight increase in SNR appears, alongside with wiggles in the mass ratio dependence. We expect these wiggles to be an artefact due to our focus on the late inspiral+merger only.
Depending on the value of $Q$, we probably cut unevenly the HM contribution, causing wiggles in SNR. Taking a longer waveform into account seems to smooth out this effect. Conversely, the memory-only part is less affected by these features, but benefits from a slight increase in SNR due to the HM contribution in Eq.~\eqref{eq:BMS_BalanceLaw}.

\begin{figure}[H]
    \centering
    \includegraphics[width=0.9\linewidth]{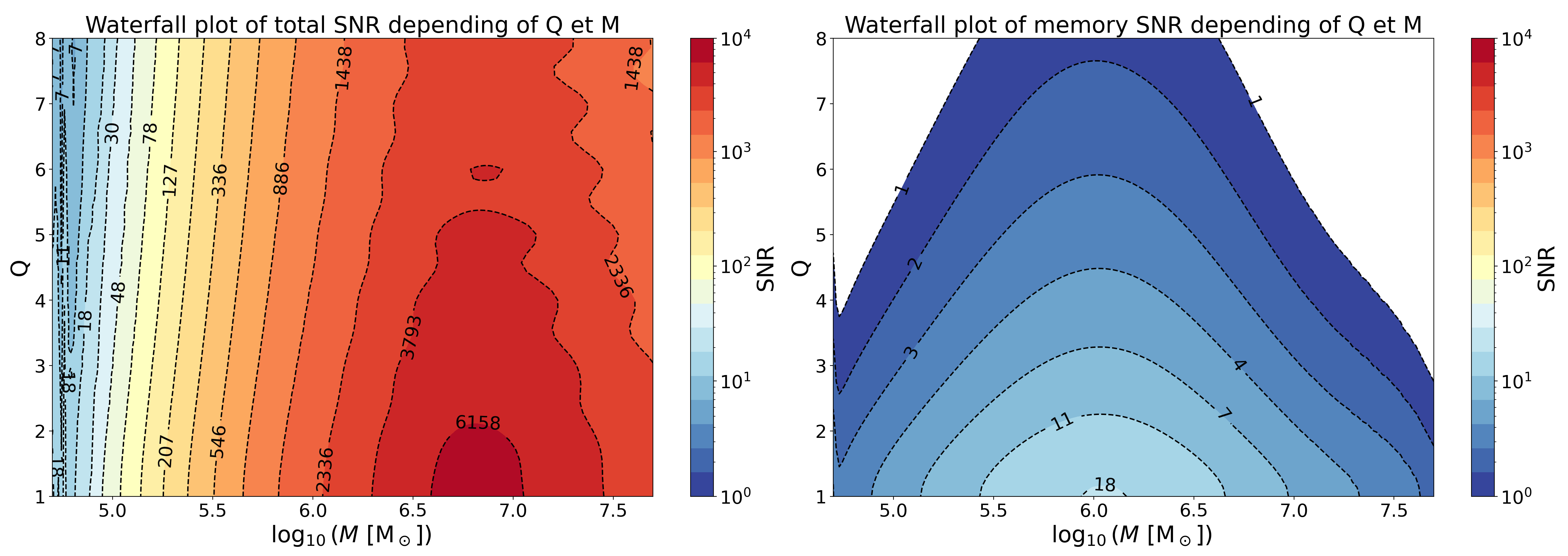}
    \caption{Total SNR (left) and SNR of the memory (right) as a function of the total mass $M$ and the mass ratio $Q$. Here we used the {\tt SEOBNRv5HM} waveform with all the previously cited modes. The other parameters used are the same as in Fig.~\ref{fig:WaterfallPlotsSurrogate}}
    \label{fig:WaterfallPlotsSEOBNR}
\end{figure}

Fig.\ref{fig:WaterfallPlotsSurrogateWithWDNoise} shows the impact of taking into account the galactic confusion noise (estimation for 4-years).

\begin{figure}[H]
    \centering
    \includegraphics[width=0.9\linewidth]{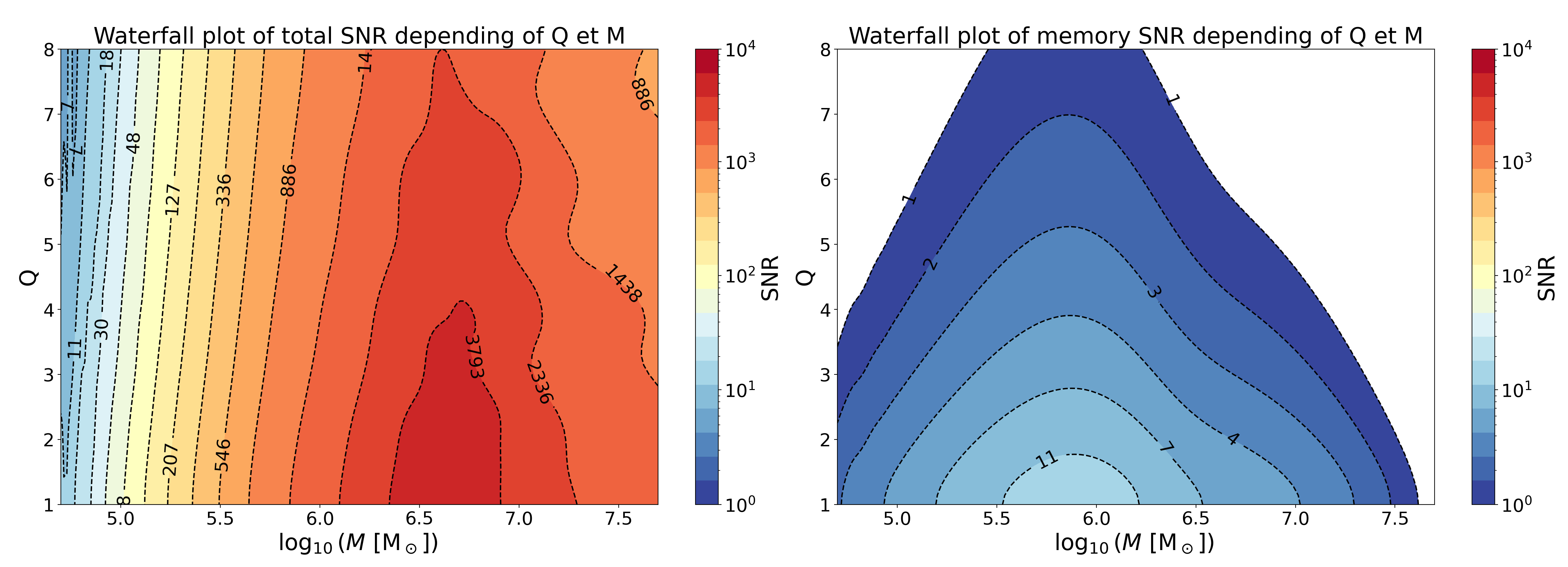}
    \caption{Total SNR (left) and SNR of the memory (right) depending on the total mass $M$ and the mass ratio $Q$. Here we used the {\tt NRHybSur3dq8\_CCE} waveform, including HMs, and add the simulated PSD of 4-years observation of the galactic confusion noise. This reduces the SNR of both total and memory signal for some total masses. The other parameters used are the same as in Fig.~\ref{fig:WaterfallPlotsSurrogate}}
    \label{fig:WaterfallPlotsSurrogateWithWDNoise}
\end{figure}

\newpage
\section{Waterfall plot in mass - redshift space}

Waterfall plots can also be used to show relation between other parameters. One of the most common type are the mass against redshift waterfall plots. A waterfall of this kind is show Fig.\ref{fig:WaterfallPlotRedshift} with its associated detectability plot.

\begin{figure}[H]
    \centering
    \includegraphics[width=0.49\linewidth]{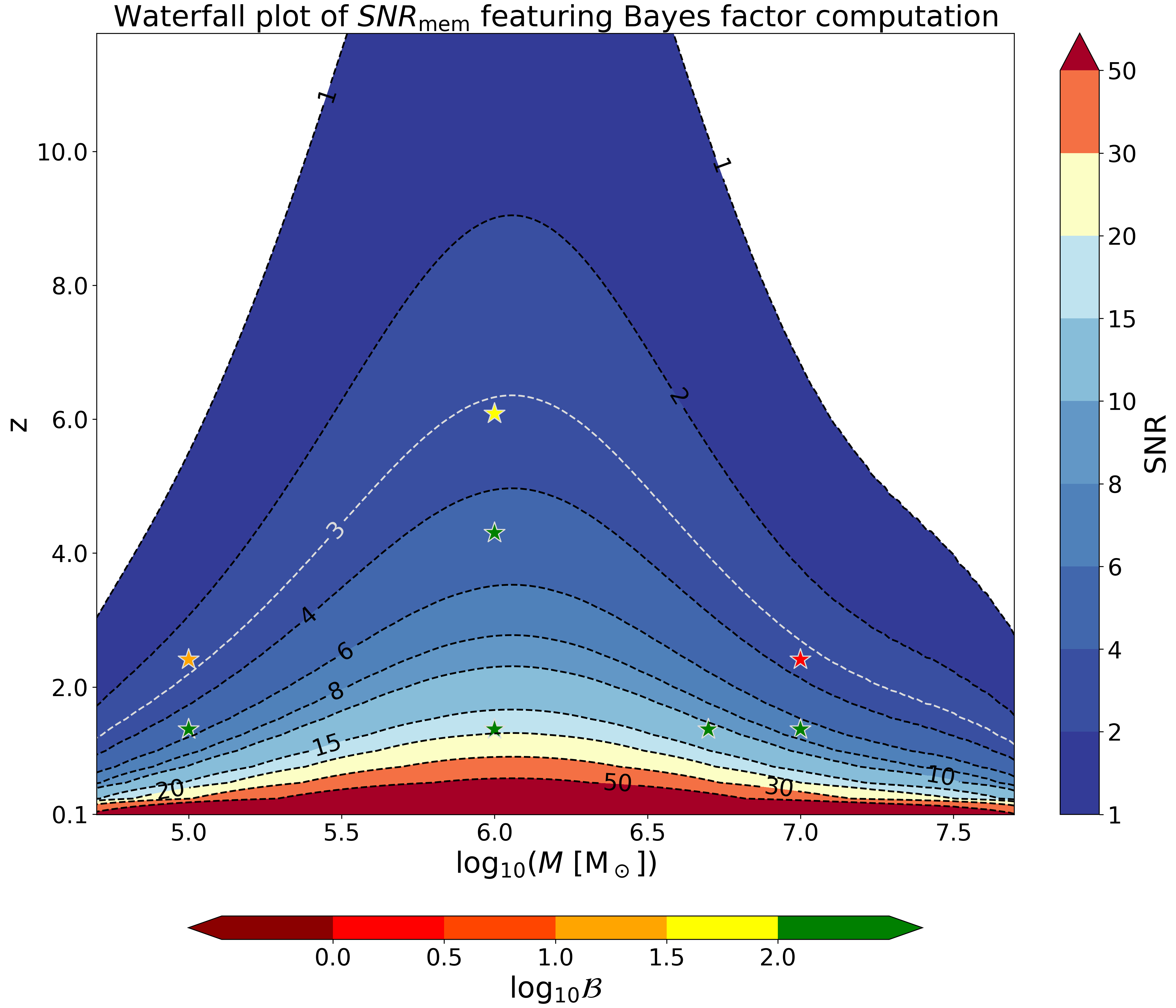}
    \includegraphics[width=0.49\linewidth]{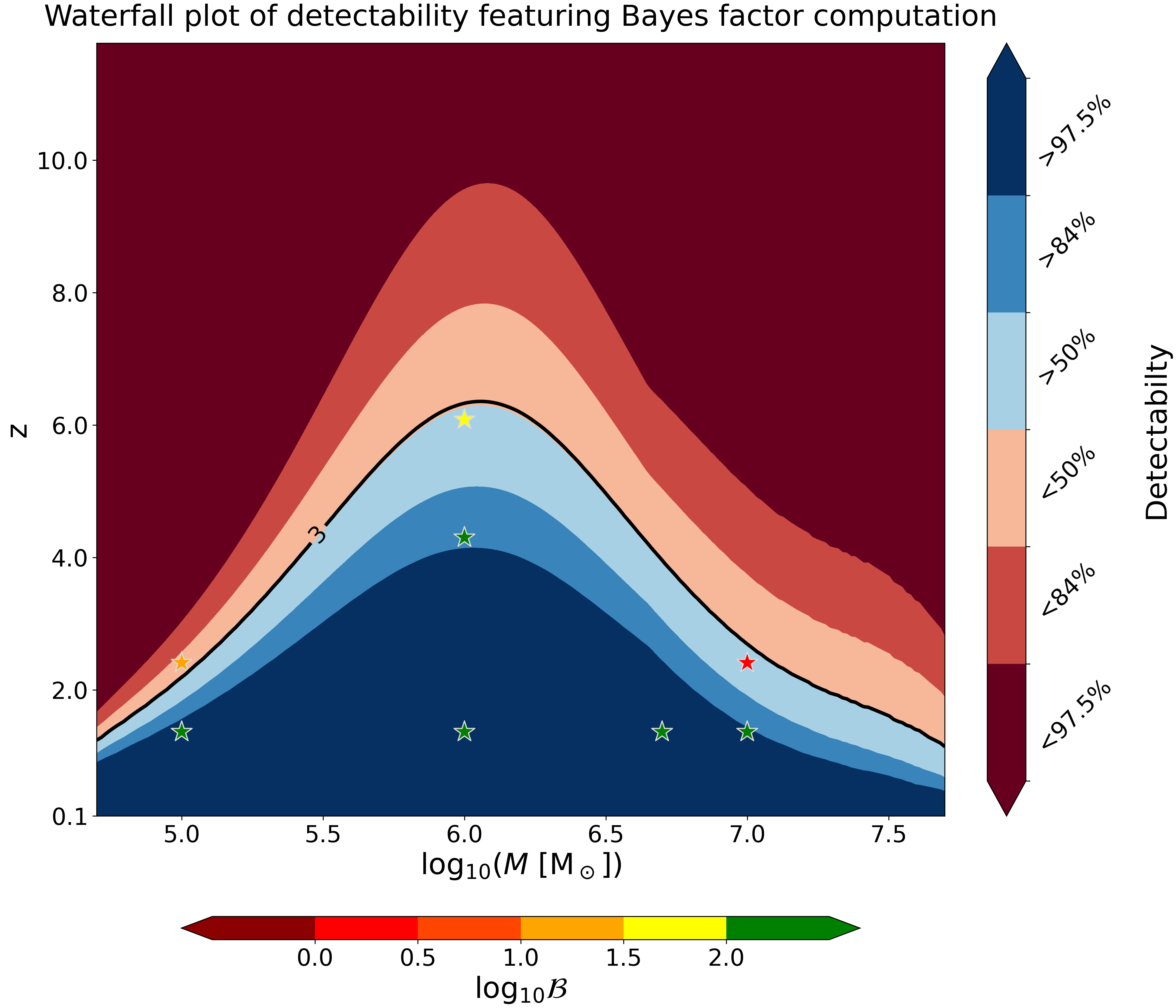}
    \caption{SNR of the memory (left) and associated detectability estimation (right) depending on the total mass $M$ (in the detector frame) and the redshift $z$. Here we used the {\tt NRHybSur3dq8\_CCE} waveform, including HM for the $\SNRmem$ computation (and the resulting detectability estimation). The Bayes factor computation stars however results from runs without HM. The difference on the favourable to detection is small between the two cases, but it can explain why some points near the $\SNRmem$ limit appear less favourable than expected. The parameters used are the same as in Fig.~\ref{fig:WaterfallPlotsSurrogate} except that $d_{\mathrm{L}}$ varies and $Q = 1$.}
    \label{fig:WaterfallPlotRedshift}
\end{figure}

\newpage
\section{Complete cornerplots}

Fig.~\ref{fig:LowSNRDoubleCornerClear} and Fig.~\ref{fig:LowSNRDoubleCornerDegenerate} show the complete cornerplots shown in Fig.~\ref{fig:DoubleDoublecornerplots}. The SNR indicated are the standard SNR, described in Eq.~\eqref{eq:SNR_formula}, without consideration for the noise realisation. For both case here, we also checked the $\SNRtot^{\textrm{mes}}$ and $\SNRmem^{\textrm{mes}}$ taking into account the noise realisation. This is obtained using eq.~\eqref{eq:Mesured_SNR}, considering $d$ as the noisy data, $m$ the noiseless model, and $\alpha$ the TDI-channel, as in eq.~\eqref{eq:SNR_formula}.
\begin{equation}
    \rho_{\alpha}^{\textrm{mes}} = \frac{\left< d | m \right>_\alpha}{\sqrt{\left< m | m \right>}_\alpha}
    \label{eq:Mesured_SNR}
\end{equation}
We observed no substantial difference between the two cases and the noiseless value.

\begin{figure}[H]
    \centering
    \includegraphics[width=0.9\linewidth]{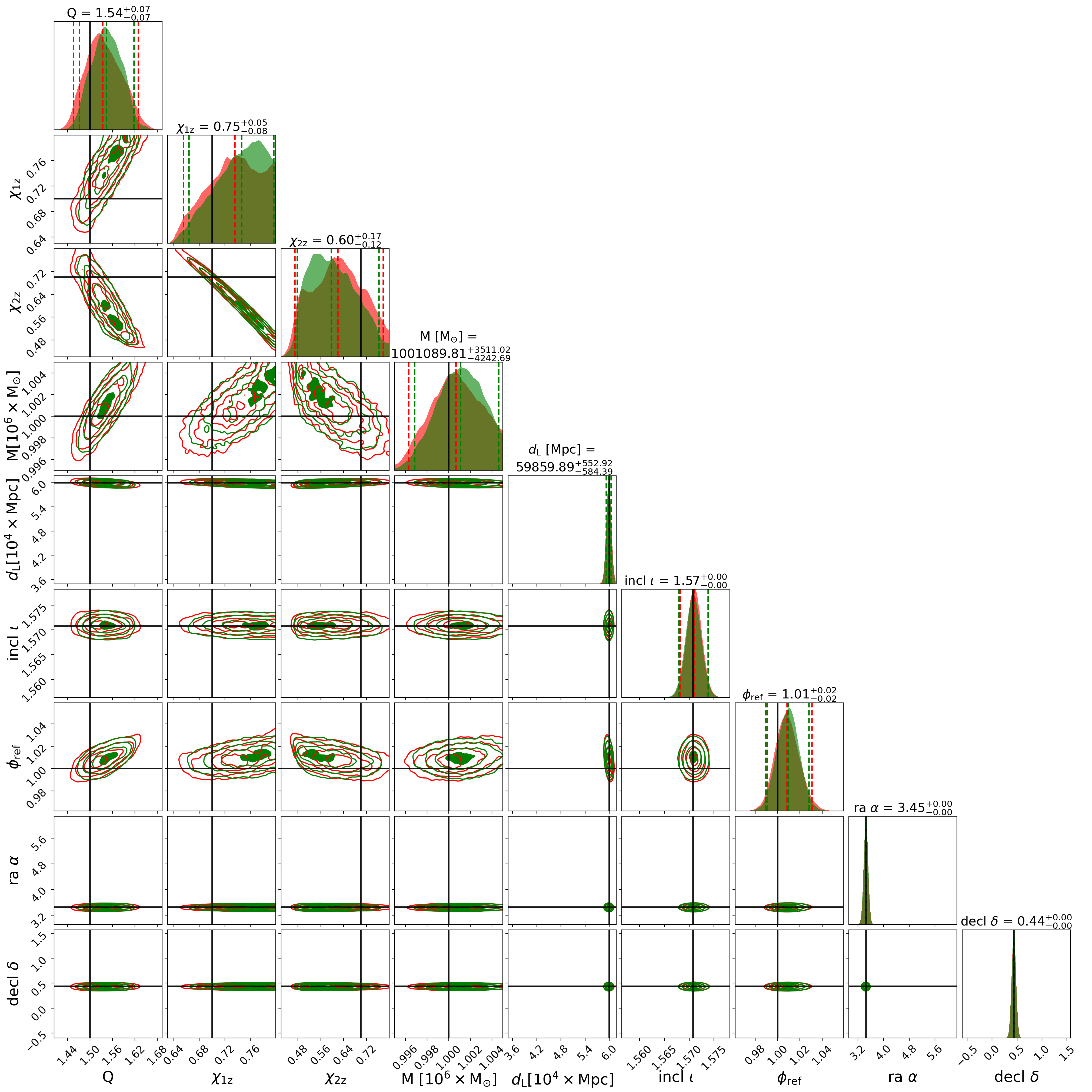}
    \caption{Cornerplot showing parameters estimation using a model with memory (green) and without (red). The values and uncertainties for parameters indicated on top of the distribution correspond to the memory model. Here we used the {\tt NRHybSur3dq8\_CCE} waveform. The associated $SNR$ values are $\SNRtot = 344$ and $\SNRmem = 6$. The injection parameters here are: $Q = 1.5$, $\chi_{\mathrm{1z}} = \chi_{\mathrm{2z}}=0.7$, $M=10^6 ~\mathrm{M_\odot}$, $d_{\mathrm{L}} = 6 \times 10^4 ~\mathrm{Mpc}$, $\iota = \pi/2$, $\varphi_{\mathrm{ref}} = 1$, $\psi = 0$, $\alpha = 3.45$, $\delta = 0.44$. This cornerplot is done with the same parameters as in Fig.~\ref{fig:LowSNRDoubleCornerDegenerate} but using a different noise realization.}
    \label{fig:LowSNRDoubleCornerClear}
\end{figure}

\begin{figure}[H]
    \centering
    \includegraphics[width=1\linewidth]{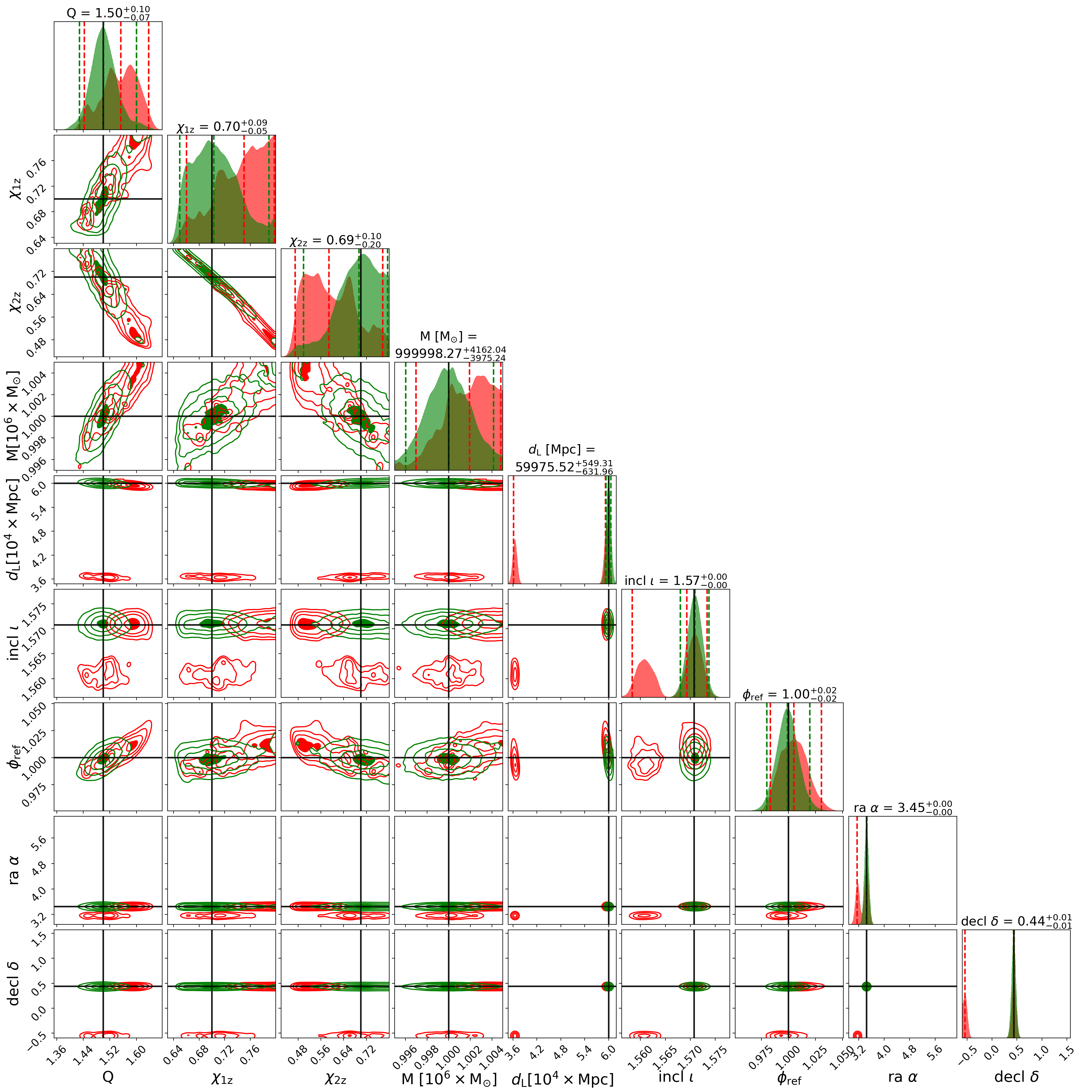}
    \caption{Cornerplot showing parameters estimation using a model with memory (green) and without (red). The values and uncertainties for parameters indicated on top of the distribution correspond to the memory model. Here we used the {\tt NRHybSur3dq8\_CCE} waveform. The associated SNR values are $\SNRtot = 344$ and $\SNRmem = 6$. The injection parameters here are: $Q = 1.5$, $\chi_{\mathrm{1z}} = \chi_{\mathrm{2z}}=0.7$, $M=10^6 ~\mathrm{M_\odot}$, $d_{\mathrm{L}} = 6 \times 10^4 ~\mathrm{Mpc}$, $\iota = \pi/2$, $\varphi_{\mathrm{ref}} = 1$, $\psi = 0$, $\alpha = 3.45$, $\delta = 0.44$. This cornerplot is done with the same parameters as in Fig.~\ref{fig:LowSNRDoubleCornerClear} but using a different noise realization.}
    \label{fig:LowSNRDoubleCornerDegenerate}
\end{figure}

\newpage
\section{Comparison between oscillatory and memory component of the $(2,0)$ mode}
\makeatletter\def\@currentlabel{Appendix D}\makeatother
\label{app:20oscill}

Let us consider a case where we use the {\tt NRHybSur3dq8\_CCE} waveform with the $(2,2)$ mode and the full $(2,0)$ mode. As this waveform natively includes memory (type $b$ waveform, see Section~\ref{sec:MemLISA}), the $(2,0)$ mode includes the memory component and an oscillatory component. Recent papers, such as~\cite{Rossello_2025}, show the effect of including the full $(2,0)$ mode on the parameter estimation, leading to consequent biases when the $(2,0)$ is not accounted for. Here, we aim to show that one of our main results \textemdash i.e., most of the time, memory does not significantly affect the parameters estimation \textemdash is compatible with the results shown in~\cite{Rossello_2025}.
Indeed, even if the $(2,0)$ oscillatory component may look negligible compared to the $(2,0)$ memory component, it is far less damped during the response and post-treatment process. Fig.~\ref{fig:Comparing20Components} shows the time-domain waveform compared to the time-domain TDI-A channel.

\begin{figure}[H]
    \centering
    \includegraphics[width=0.43\linewidth]{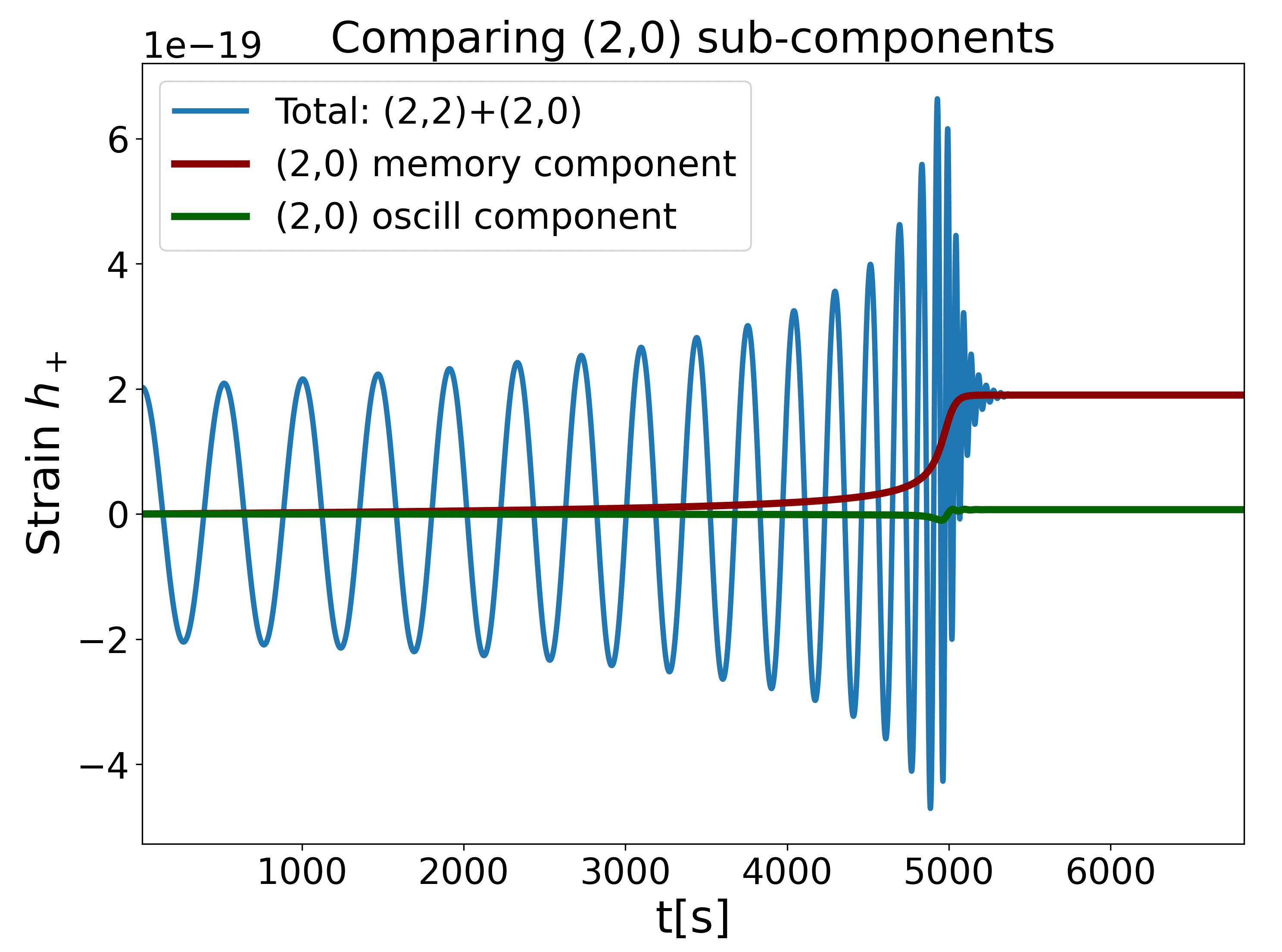}
    \includegraphics[width=0.53\linewidth]{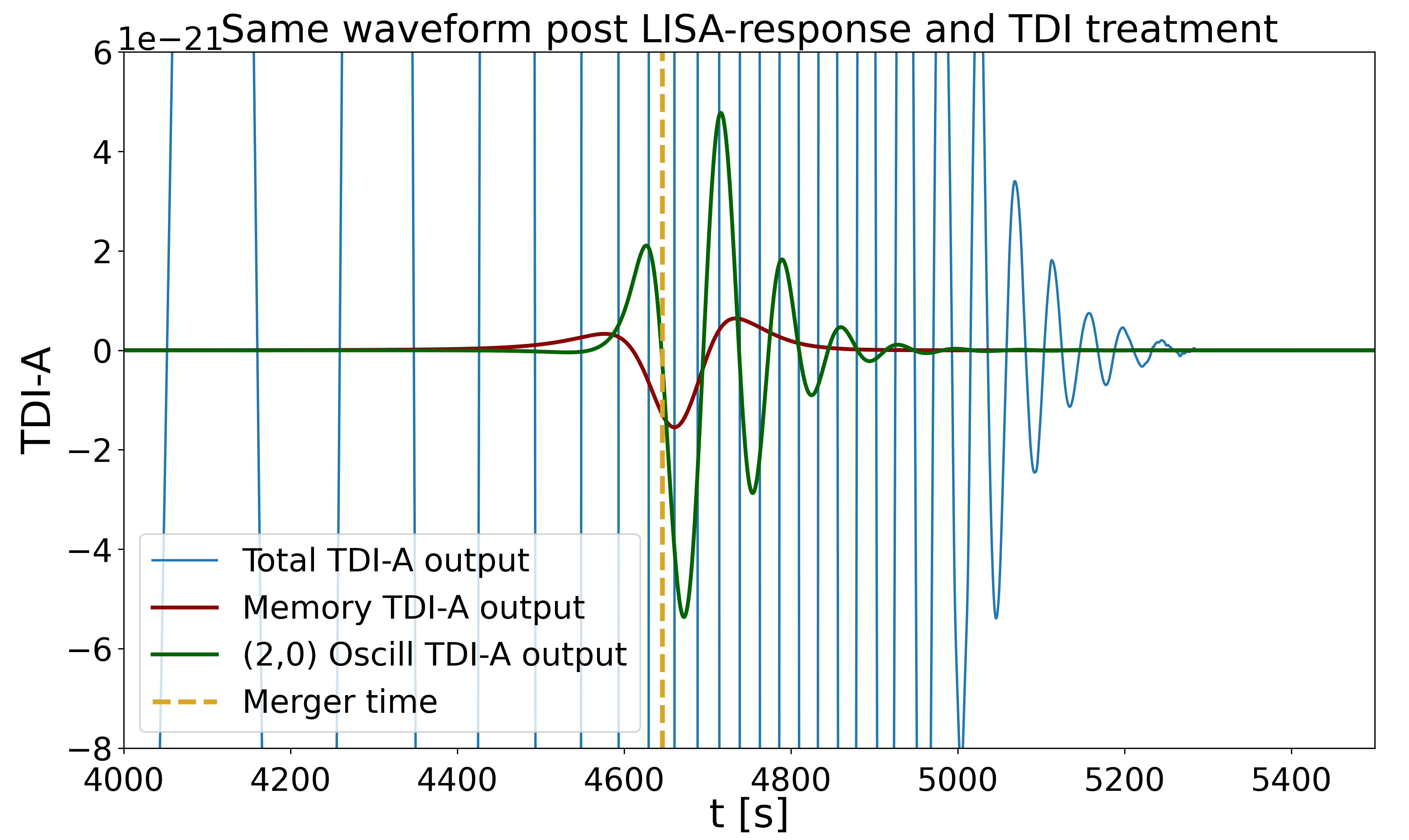}
    \caption{Comparison between the two components of the complete (2,0) mode. In the time-domain waveform (left), we can see that the oscillatory part of the (2,0) looks negligible compared to the (2,0) memory component. However the TDI-A time domain channel (right) shows that the memory component is way more suppressed by LISA response and TDI post-treatment.
    Parameters: $Q=1.5$, $\chi_{\mathrm{1z}} = \chi_{\mathrm{2z}} = 0.6$, $M = 10^6 ~\mathrm{M_\odot}$, $d_{\mathrm{L}} = 10^4 ~\mathrm{Mpc}$, $\iota = \pi/2$, $\varphi_{\mathrm{ref}} = 0$, $\psi=0$, $\alpha = 0.74$, $\delta = 0.28$.}
    \label{fig:Comparing20Components}
\end{figure}

Running a parameter estimation with this scenario gives us the following corner plot, see Fig.~\ref{fig:Cornerplot20}, where we can compare the effect of taking or not into account the (2,0) oscillatory component. The mock data used in this case contains both (2,2) and full (2,0) modes. The parameter estimation is done with a (2,2) and full (2,0) model (in blue), and with a model including only the (2,2) and the memory component of the (2,0). As a result, we can observe clear biases in parameter estimation when we neglect the oscillatory component of the (2,0). This is coherent with the results from~\cite{Rossello_2025}.

\begin{figure}[H]
    \centering
    \includegraphics[width=1\linewidth]{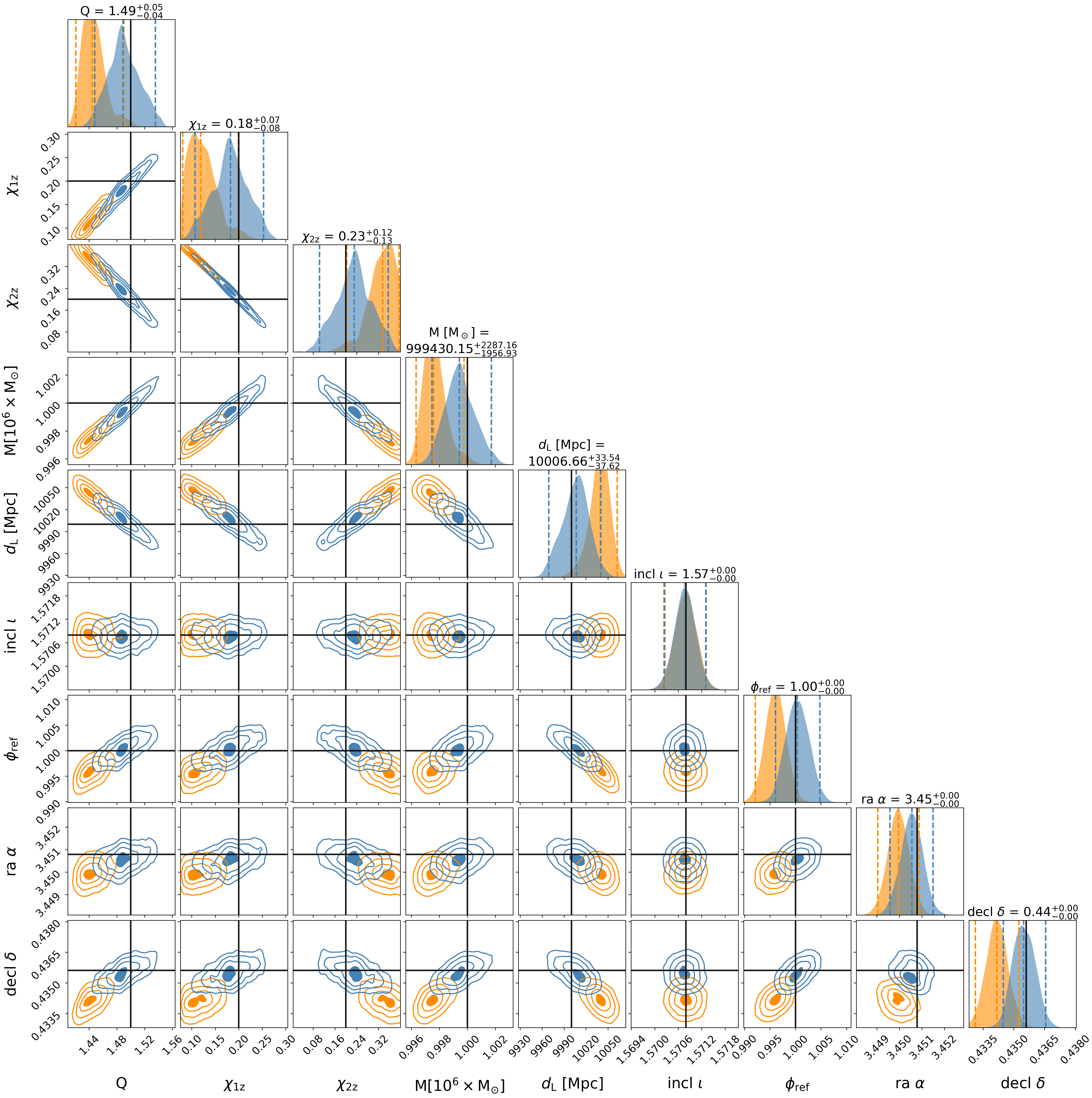}
    \caption{Cornerplot showing parameters estimation using a model with the (2,2)-mode and the full (2,0)-mode (blue) compared to a model neglecting the oscillatory component of the (2,0)-mode (orange). The data used as an input are built with the (2,2)-mode and the full (2,0). {\tt NRHybSur3dq8\_CCE} is used. Injection parameters are located with the black lines : $Q=1.5$, $\chi_{\mathrm{1z}} = \chi_{\mathrm{2z}} = 0.2$, $M = 10^6 ~\mathrm{M_\odot}$, $d_{\mathrm{L}} = 10^4 ~\mathrm{Mpc}$, $\iota = \pi/2$, $\varphi_{\mathrm{ref}} = 1$, $\psi = 0$, $\alpha = 3.45$, $\delta = 0.44$.}
    \label{fig:Cornerplot20}
\end{figure}

\newpage
\section{Complementary results on estimation with catalogue}
Fig.~\ref{fig:Barausse_10yrs_MemorySeenSources} shows the distribution of the number of sources with visible memory in 10-years realizations. The Fig.~\ref{fig:Barausse_10yrs_ProbaOfMax} shows the probability of having a certain maximum value in $\SNRmem$ for both 4-years and 10-years realizations. 

\begin{table}[H]
 \begin{ruledtabular}
    \begin{tabular}{cccccccccccc}
        \multicolumn{3}{c}{Probability [\%] of having at least}  & \multirow{2}{*}{5} & \multirow{2}{*}{10} & \multirow{2}{*}{15} & \multirow{2}{*}{20} & \multirow{2}{*}{25} & \multirow{2}{*}{30} & \multirow{2}{*}{40} & \multirow{2}{*}{50} & \multirow{2}{*}{60}\\
        \multicolumn{3}{c}{one event with $\SNRmem$ above} & & & & & & & & & \\
        \hline
         \multirow{4}{*}{SN} & Short & Light & 0 & 0 & 0 & 0 & 0 & 0 & 0 & 0 & 0 \\ 
                            \cline{3-12}
                            & Delays & Heavy & 100 & 100 & 97.4 & 87.2 & 70.8 & 55.8 & 35.8 & 23.7 & 14.9\\
                            \cline{2-12}
                            & \multirow{2}{*}{Delays} & Light & 0 & 0 & 0 & 0 & 0 & 0 & 0 & 0 & 0\\
                            \cline{3-12}
                            &   & Heavy & 100 & 97.5 & 84.8 & 68.1 & 51.6 & 40.7 & 23.0 & 14.1 & 8.90 \\
        \hline
         \multirow{4}{*}{NoSN} & Short & Light & 62.4 & 25.1 & 8.20 & 3.50 & 1.0 & 0.20 & 0 & 0 & 0 \\
                            \cline{3-12}
                            & Delays & Heavy & 100 & 100 & 96.3 & 81.1 & 64.6 & 50.8 & 29.9 & 17.5 & 10.2 \\
                            \cline{2-12}
                            & \multirow{2}{*}{Delays} & Light & 58.7 & 32.4 & 16.5 & 8.80 & 6.30 & 4.90 & 3.60 & 2.30 & 1.30 \\
                            \cline{3-12}
                            &  & Heavy & 94.10 & 73.6 & 51.0 & 35.4 & 23.7 & 17.4 & 11.3 & 8.0 & 5.30
    \end{tabular}
    \caption{Table with the expectation of detection for different models, using Barausse's populations. This table is a numeric summary of~\ref{fig:Barausse_4yrs_ProbaOfMax}.}
    \label{tab:BarausseMaxSNRProba}
 \end{ruledtabular}
\end{table}

\begin{figure}[H]
    \centering
    \includegraphics[width=0.9\linewidth]{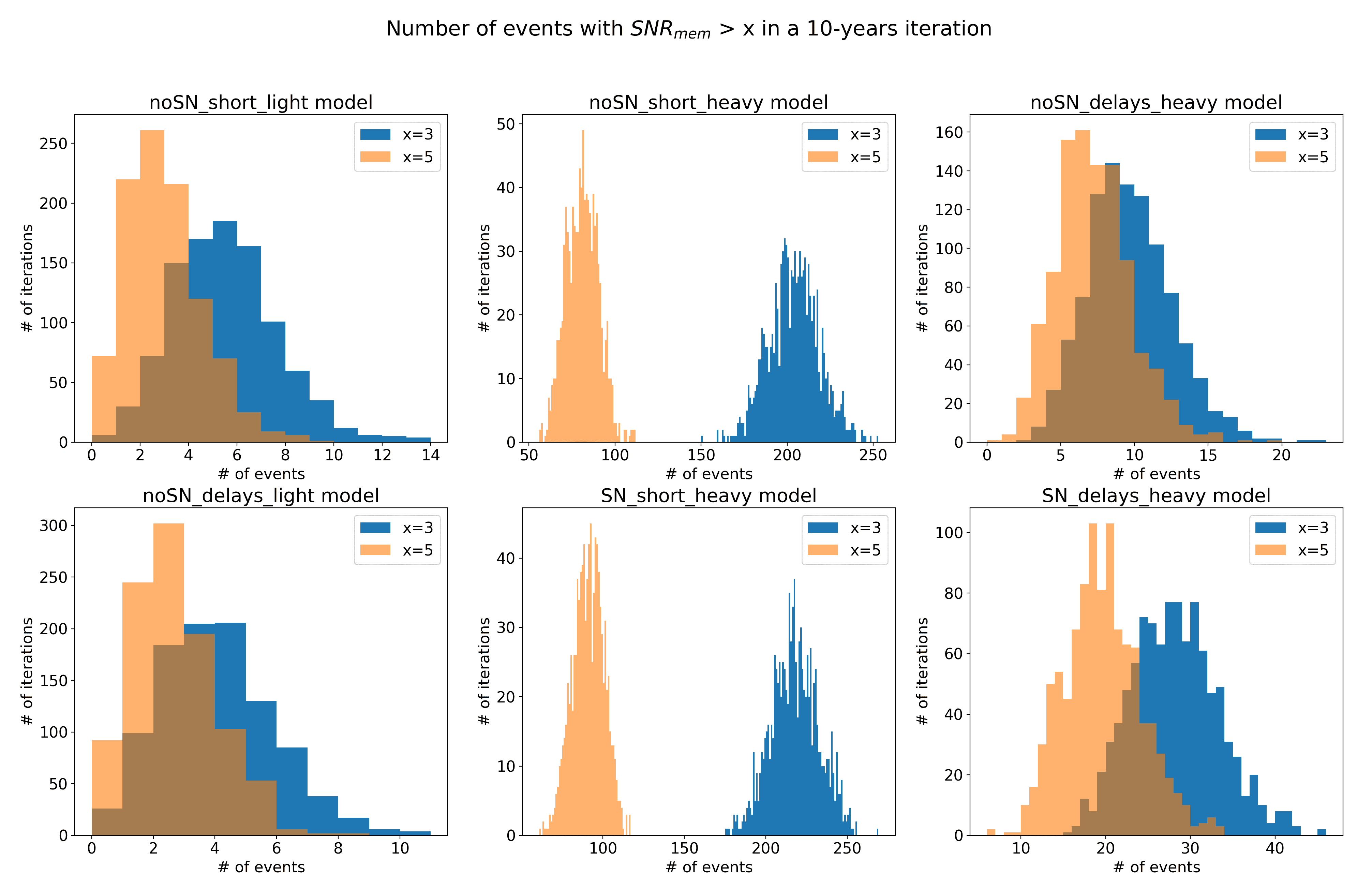}
    \caption{Histograms of the number of events such that $\SNRmem > 3$ (blue) and $\SNRmem > 5$ (orange) for the six remaining Barausse's catalogues. The bins are unitary and each models presents realizations of 10-years data.}
    \label{fig:Barausse_10yrs_MemorySeenSources}
\end{figure}

\begin{figure}[H]
    \centering
    \includegraphics[width=1\linewidth]{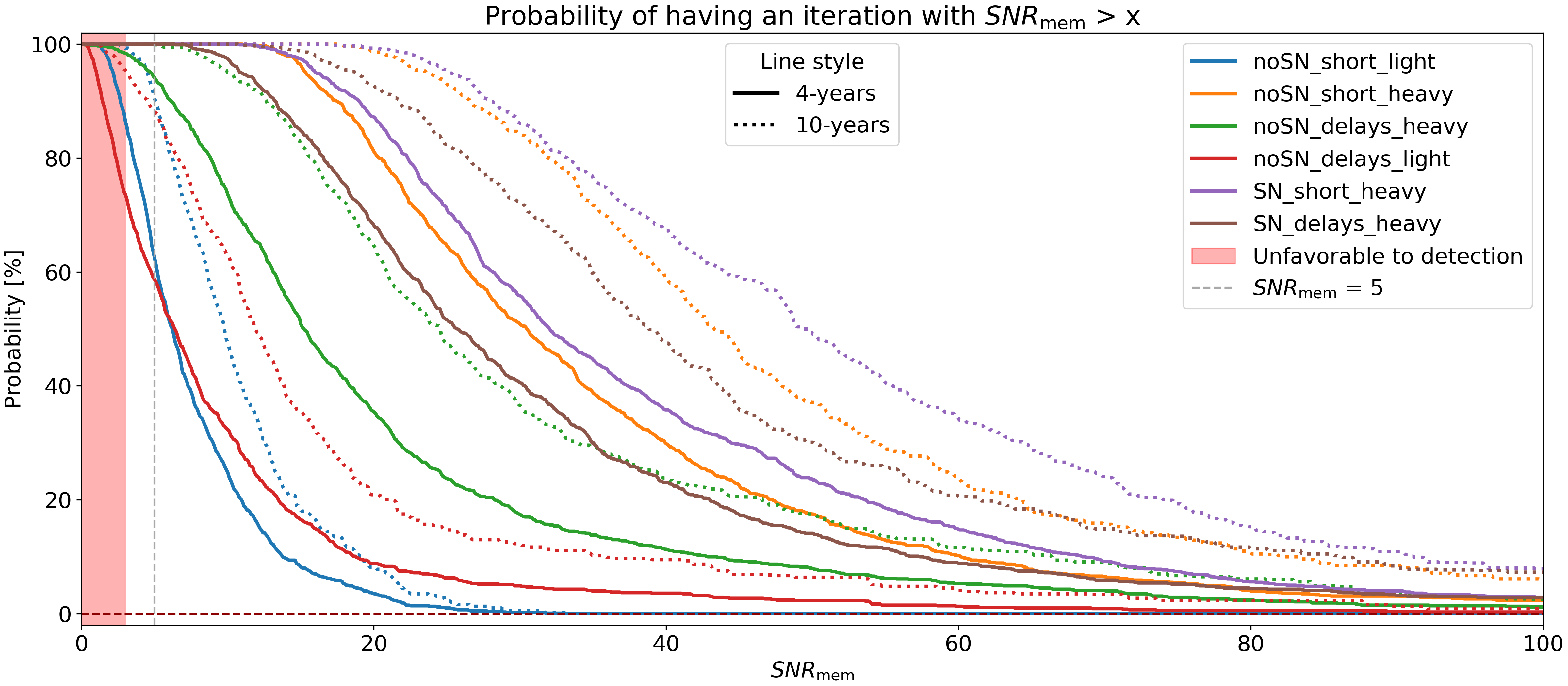}
    \caption{Probability of having an iteration with $\SNRmem$ greater than a given value (x-axis). Each color correspond to a population model from Barausse et al., 2020~\cite{Barausse_2020, Barausse_Lapi_2021}. Solid lines corresponds to 4-years iterations and dotted lines to 10-years. The red area cover the region where we are under the threshold $\SNRmem^{\textrm{thresh}} = 3$. The gray dashed line shows the value $\SNRmem = 5$ over which memory should be always detected.}
    \label{fig:Barausse_10yrs_ProbaOfMax}
\end{figure}
\end{widetext}

\newpage
$~$
\newpage
$~$
\newpage
\bibliographystyle{apsrev}
\bibliography{refs}
\end{document}